\documentclass[11pt,letterpaper]{article}

\usepackage{jcappub}

\usepackage{amsmath}
\usepackage{epsfig,multicol,bbm}
\usepackage{graphicx}
\usepackage{bm}
\usepackage{epsfig}
\usepackage{url}
\usepackage{slashed}

\usepackage{subfig}

\usepackage{multirow}

\renewcommand{\thesection}{\arabic{section}}
\renewcommand{\emph}[1]{{\it #1}}

\newcommand{\be}{\begin{equation}}
\newcommand{\ee}{\end{equation}}

\newcommand{\GeV}{\text{GeV}}
\newcommand{\TeV}{\text{TeV}}

\newcommand{\fb}{\text{fb}}
\newcommand{\pb}{\text{pb}}

\newcommand{\Q}{{\boldsymbol \Psi}}
\newcommand{\Qc}{{\boldsymbol{\Psi^c}}}
\newcommand{\Qscalar}{\widetilde{\psi}}
\newcommand{\QscalarOne}{\widetilde{\psi}_1}
\newcommand{\QscalarTwo}{\widetilde{\psi}_2}
\newcommand{\Qfermion}{\psi}
\newcommand{\Qcscalar}{\widetilde{\psi}^c}
\newcommand{\Qcscalarstar}{\widetilde{\psi}^{c\,\dag}}
\newcommand{\Qcfermion}{\psi^c}
\newcommand{\Qfermiondag}{\overline{\psi}}
\newcommand{\Qcfermiondag}{\overline{\psi^{c}}}
\newcommand{\QDirac}{\Psi}
\newcommand{\scalars}{\widetilde{\Psi}}
\newcommand{\bino}{\widetilde{\chi}}

\newcommand{\etaB}{\eta_B}
\newcommand{\Yeq}{Y^{\rm eq}}

\newcommand{\Eqs}[2]{Eqs.~\eqref{#1} and \eqref{#2}}
\newcommand{\Eq}[1]{Eq.~\eqref{#1}}
\newcommand{\Tab}[1]{Table~\ref{#1}}
\newcommand{\Sec}[1]{Sec.~\ref{#1}}
\newcommand{\Fig}[1]{Fig.~\ref{#1}}
\newcommand{\Figs}[2]{Figs.~\ref{#1} and \ref{#2}}
\newcommand{\App}[1]{App.~\ref{#1}}
\newcommand{\Ref}[1]{Ref.~\cite{#1}}
\newcommand{\Refs}[1]{Refs.~\cite{#1}}

\def\sigmabar{\overline\sigma}

\def\psibar{{\bar{\psi}}}

\title{Dark Matter Assimilation into the Baryon Asymmetry}

\author{Francesco D'Eramo,}
\author{Lin Fei,}
\author{and Jesse Thaler}  

\affiliation{Center for Theoretical Physics, Massachusetts Institute of Technology, \\
77 Massachusetts Avenue, Cambridge, MA 02139, U.S.A.} 

\emailAdd{fderamo@mit.edu}
\emailAdd{lfei@mit.edu}
\emailAdd{jthaler@mit.edu} 
 
\abstract{Pure singlets are typically disfavored as dark matter candidates, since they generically have a thermal relic abundance larger than the observed value.  In this paper, we propose a new dark matter mechanism called ``assimilation'', which takes advantage of the baryon asymmetry of the universe to generate the correct relic abundance of singlet dark matter.  Through assimilation, dark matter itself is efficiently destroyed, but dark matter number is stored in new quasi-stable heavy states which carry the baryon asymmetry.  The subsequent annihilation and late-time decay of these heavy states yields (symmetric) dark matter as well as (asymmetric) standard model baryons. We study in detail the case of pure bino dark matter by augmenting the minimal supersymmetric standard model with vector-like chiral multiplets.  In the parameter range where this mechanism is effective, the LHC can discover long-lived charged particles which were responsible for assimilating dark matter.}

\begin{document}

\hfill MIT-CTP 4327

\maketitle


\section{Introduction}

The existence of dark matter (DM) is striking evidence for physics beyond the standard model (SM) \cite{Jungman:1995df,Bergstrom:2000pn,Bertone:2004pz,Ellis:2010kf,Feng:2010gw}. Stable particles with Fermi-scale masses and electroweak annihilation cross sections, known as Weakly Interacting Massive Particles (WIMPs), are prime DM candidates since they have the right thermal relic abundance to account for observations \cite{Lee:1977ua,Scherrer:1985zt,Kolb:1985nn,Srednicki:1988ce,Gondolo:1990dk}.  Supersymmetry (SUSY) with conserved $R$-parity naturally provides WIMP DM if the lightest supersymmetric particle (LSP) is a neutralino \cite{Goldberg:1983nd,Krauss:1983ik,Ellis:1983ew,Bottino:1993zx,ArkaniHamed:2006mb}.  This can occur in the Minimal Supersymmetric Standard Model (MSSM) as well as in various generalizations.

Depending on the soft mass spectrum, it is often the case that the LSP is a nearly pure bino eigenstate.  The bino is a gauge singlet, and its annihilation through $p$-wave-suppressed $t$-channel sfermion exchange is too small to yield the correct DM relic density.  For this reason, a pure bino is typically excluded by overclosure, and mixed neutralino DM is often assumed in the MSSM.  There are of course ways to accommodate bino DM.  The bino might be almost degenerate in mass with the lightest sfermion such that thermal production can be dominated by co-annihilation \cite{Griest:1990kh,Ellis:1998kh,Ellis:1999mm}.  Alternatively, one can choose a reheat temperature lower than the freezeout temperature to achieve the correct bino relic density \cite{Giudice:2000ex}.  Besides the MSSM bino, SM gauge singlets are well-motivated DM candidates, and the correct thermal abundance can arise if the singlet interacts with portal fields that provide additional annihilation channels \cite{Burgess:2000yq,Kim:2008pp,Patt:2006fw,MarchRussell:2008yu,Nomura:2008ru}, or if the singlet mixes with ``higgsino-like'' states \cite{Mahbubani:2005pt,D'Eramo:2007ga,Enberg:2007rp,Cohen:2011ec}.

In this paper, we propose an alternative mechanism to achieve the correct thermal abundance of singlet DM, by taking advantage of matter/antimatter asymmetries.  We introduce a new DM process called ``assimilation'', which takes the schematic form
\be
\bino \, \QDirac \; \rightarrow \scalars \, \phi \ .
\label{eq:DMA}
\ee 
Here, $\bino$ is singlet DM, $\QDirac$ and $\scalars$ are new quasi-stable heavy states which carry an asymmetry, and $\phi$ is a SM field.  The tildes in $\bino$ and $\scalars$ denote a multiplicatively conserved DM number, such as $R$-parity in SUSY theories.  In the presence of a $\QDirac$ asymmetry, the process in \Eq{eq:DMA} is very effective at destroying singlet particles $\bino$.   In the process, DM number (e.g.\ $R$-parity) is conserved and stored in $\scalars$.  In this way, DM is ``assimilated'' into the heavier state $\scalars$.

The reason an asymmetry is necessary is that $\QDirac$ has to be sufficiently dense at temperatures of the order of the DM mass for \Eq{eq:DMA} to be relevant.  Without an asymmetry, $\QDirac$ and $\scalars$ are quickly depleted by the crossed version of \Eq{eq:DMA}.  Motivated by the baryon/antibaryon asymmetry, we will focus on the case where $\QDirac$ and $\scalars$ are charged under $U(1)_B$ and carry the full baryon asymmetry at early times.  We will not give an explicit model of $\QDirac$-ogenesis in this paper, though certain standard baryogenesis mechanisms can be adapted to generate a $\QDirac$ asymmetry.  Note that unlike asymmetric DM \cite{Nussinov:1985xr,Kaplan:1991ah,Barr:1990ca,Barr:1991qn,Dodelson:1991iv,Fujii:2002aj,Farrar:2004qy,Hooper:2004dc,Kitano:2004sv,Farrar:2005zd,Kitano:2008tk,Kaplan:2009ag}, DM itself does not have an asymmetry.  Unlike in co-annihilation \cite{Griest:1990kh}, $\QDirac$ and $\scalars$ are sufficiently long-lived that they can be considered stable during DM freezeout.\footnote{Quasi-stability of $\QDirac$ and $\scalars$ requires that the triangle inequality among the particle masses $m_{\scalars} < m_{\QDirac} + m_{\Psi}$ is satisfied (as well as its crossed versions). In the limiting case where $\QDirac$ and $\scalars$ are absolutely stable, the reaction in \Eq{eq:DMA} is an example of semi-annihilation \cite{D'Eramo:2010ep}.}  

Since $\scalars$ are not gauge singlets, they typically have large annihilation cross sections, especially in SUSY theories.  Assuming that $\scalars$ is heavier than $\QDirac$, the process
\be
\scalars\scalars \rightarrow \QDirac \QDirac 
\ee
efficiently depletes DM number until $\scalars$ freezes out.  Analogous to the super-WIMP scenario \cite{Feng:2003xh,Feng:2003uy}, at late times, the decay of $\scalars$ produces singlet particles $\bino$ which compromise DM in our universe today.  Since $\QDirac$ and $\scalars$ carry an asymmetry, it is natural (but not necessary) for their decay products to generate the baryon asymmetry, through
\be
\QDirac \rightarrow \text{quarks} \ , \qquad \scalars \rightarrow \bino + \text{quarks} \ . 
\ee
In this way, DM and baryons can arise from the late decay of a common parent particle, leaving a baryon asymmetry but symmetric DM.

This mechanism of DM assimilation is general and can be implemented in many singlet DM models.  In this work, we focus on the SUSY bino case for concreteness.  We consider a minimal extension of the MSSM, where we only add a heavy vector-like chiral multiplet $\Q/\Qc$.  The only new parameters with respect to the MSSM alone are the superpotential mass and soft terms for these new states.  The new $R$-odd degrees of freedom destroy and ``assimilate'' bino DM in the early universe through the process in \Eq{eq:DMA}.  As long as $\Q/\Qc$ have the right asymmetry and the correct quantum numbers, their late time decays can be responsible for generating the baryon asymmetry.  We find that this mechanism works best when $\Q/\Qc$ are at the TeV scale, implying that the LHC is able to probe this physics by looking for quasi-stable charged particles.  
 
The remainder of this paper is structured as follows.  In \Sec{sec:model}, we introduce a simple model of DM assimilation and mention possible generalizations.  In \Sec{sec:Boltz}, we study bino assimilation in the early universe, showing both semi-analytical and numerical results, and leaving calculational details to the appendices.  We discuss the experimental implications for DM searches and collider measurements in \Sec{sec:exp}, and conclude in \Sec{sec:con}.

\section{A Simple Model of Dark Matter Assimilation}
\label{sec:model}

In this section, we present a simple model of DM assimilation involving new quasi-stable particles.  These new particles carry a matter/antimatter asymmetry which we identify with SM baryon number, and their late time decay generates both DM and baryons.  In \Sec{sec:generalizations}, we describe a few generalizations of this model.

\subsection{Field Content and Lagrangian}
\label{eq:lagrangian}

\begin{table}
\begin{center}
\begin{tabular}{|c||c|c|}
\hline
& $\left(SU(3)_C, SU(2)_L, U(1)_Y\right)$ & $U(1)_B$ \\ \hline\hline
$\Q$ & $({\bf 1}, {\bf 1}, +1)$  & $ + 1$ \\
$\Qc$ & $({\bf 1}, {\bf 1}, - 1)$  & $ - 1$ \\ \hline
\end{tabular}
\end{center}
\caption{Additional superfield content compared to the MSSM for a simple model of DM assimilation.  Here, we consider multiplets that only carry hypercharge, with generalizations shown in \Tab{tab:altFields}.}
\label{tab:fields}
\end{table}

In our model, the MSSM is augmented by vector-like superfields $\Q/\Qc$ that carry hypercharge and baryon number.  The quantum numbers of $\Q/\Qc$ are shown in \Tab{tab:fields}, with alternatives discussed in \Sec{sec:generalizations}.  The leading superpotential terms consistent with the symmetries are
\be
\label{eq:superpotential}
W = W_{{\rm MSSM}} + M \Q \Qc \ .
\ee
We will work in the limit of a (nearly) pure bino, so the only MSSM multiplet relevant for our discussion is the hypercharge gauge multiplet, containing the hypercharge gauge boson $B_\mu \equiv \cos \theta_W \gamma_\mu + \sin \theta_W Z_\mu$ with coupling $\alpha_Y \equiv g_Y^2/(4\pi)$, and the bino $\bino$ with soft mass $m_{\bino}$.  We assume $R$-parity conservation throughout this paper, and tildes refer to $R$-odd fields.   This superpotential has a continuous $U(1)_B$ symmetry, under which $\Q$ and $\Qc$ have charges $\pm 1$, and we identify $U(1)_B$ with SM baryon number through decay operators shown in \Sec{sec:latedecay}.  

\begin{table}
\begin{center}
\begin{tabular}{|c||c|c|c|c|c|}
\hline
 & {Type} & {Mass} & $(SU(3)_C, SU(2)_L, U(1)_Y)$ &  $U(1)_B$ & $R$-parity \\
\hline\hline
$\QscalarOne$ & Complex Scalar & $M +  \frac{m_{{\rm soft}}^2 + b_M}{2 M} $  & $({\bf 1},{\bf 1},+1)$ & $+1$ & $-$ \\
$\QscalarTwo$ & Complex Scalar & $M +  \frac{m_{{\rm soft}}^2 -  b_M}{2 M} $  & $({\bf 1},{\bf 1},+1)$ & $+1$ & $-$ \\
$\QDirac$ & Dirac Fermion & $M $  & $({\bf 1},{\bf 1},+1)$ & $+1$ & $+$ \\ \hline
\end{tabular}
\end{center}
\caption{Component fields and their properties for the heavy states.  The masses are given in the limit $\delta \ll M$, with $\delta$ defined in \Eq{eq:deltamass}.  For the relic abundance calculations, we will ignore the $b_M$ mass contribution for simplicity.}
\label{tab:comp}
\end{table} 

The physical degrees of freedom are summarized in \Tab{tab:comp}.  Including the relevant SUSY-breaking soft terms, the scalar masses are 
\be
\mathcal{L}_{{\rm scalar}}  \supset - (M^2 + m_{\rm soft}^2) \left(\left|\Qscalar\right|^2 + \left|\Qcscalar\right|^2 \right) - b_M \left(\Qscalar \Qcscalar +  \mathrm{h.c.} \right) \ ,
\label{eq:Lsoft}
\ee
where $\Qscalar$ ($\Qcscalar$) is the scalar component of $\Q$ ($\Qc$).  In the presence of the $b_M$ term, the scalar mass eigenstates are
\be
\QscalarOne = \frac{\Qscalar + \Qcscalarstar}{\sqrt{2}} \ , \qquad \QscalarTwo = \frac{\Qscalar - \Qcscalarstar}{\sqrt{2}} \ ,
\ee
with masses
\be
m_{1,2} = \sqrt{M^2 + m_{\rm soft}^2 \pm b_M} \ .
\ee
For simplicity, we will neglect the $b_M$ terms for the relic abundance calculation in \Sec{sec:Boltz}, but they are important when we discuss the scalar decays in \Sec{sec:latedecay}.  We will use the notation $\scalars$ to refer to both $\QscalarOne$ and $\QscalarTwo$
\be
\label{eq:simpScalarNotation}
\scalars \equiv \{\QscalarOne,\QscalarTwo\} \ , \qquad \scalars^\dagger \equiv \{\QscalarOne^\dagger,\QscalarTwo^\dagger\} \ ,
\ee
since they have the same properties up to the small mass difference.

It is convenient to combine the Weyl fermions $\Qfermion$/$\Qcfermion$ in $\Q$/$\Qc$ into a single Dirac field $\QDirac$
\be
\QDirac = \left( \begin{array}{c} \Qfermion \\ \Qcfermiondag \end{array} \right) \ , 
\ee
with mass term
\be
\mathcal{L}_{\rm fermion} \supset- M \overline{\QDirac} \QDirac \ .
\ee
Ignoring the $b_M$ term, the fermion/scalar mass splitting from the scalar soft terms is
\be
\label{eq:deltamass}
\delta \equiv m_{\rm scalar} - m_{\rm fermion} = \sqrt{M^2 + m_{{\rm soft}}^2} - M \simeq \frac{m_{{\rm soft}}^2}{2 M} \ ,
\ee
and we will assume that $\delta > 0$, such that the $R$-odd scalars $\scalars$ are heavier than the $R$-even fermions $\QDirac$.  Since we expect $m_{\bino}$ to be comparable to $m_{{\rm soft}}$, the triangle inequality $m_{\scalars} < m_{\QDirac} + m_{\Psi}$ is generically satisfied and the decay $\scalars \rightarrow \QDirac \bino$ is forbidden.

\subsection{Matter/Antimatter Asymmetry}

With a mass $M \simeq \text{few}~\TeV$, the abundance of $\QDirac$ and  $\scalars$ would be quickly depleted in the absence of some kind of primordial asymmetry.  In order to have a large enough density of $\QDirac$ and  $\scalars$ to affect DM thermal freezeout, $\QDirac$ and  $\scalars$ will have a (shared) asymmetry dictated by SM baryon number $U(1)_B$.  In particular, we assume that the \emph{full} SM asymmetry\footnote{In contrast to the standard notation, we use a comoving definition of $\eta_B$.}
\be
\etaB \equiv \frac{n_B - n_{\overline{B}}}{s} \simeq 8.6 \times 10^{-11}
\ee
is stored in $\QDirac$ and  $\scalars$ in the early universe, and transferred to SM baryons via the late decay of $\QDirac$ and  $\scalars$.  Here, $n_B$ ($n_{\overline{B}}$) is the baryon (antibaryon) number density and $s$ is the entropy density of relativistic degrees of freedom.  We do not give an explicit model for $\QDirac$-ogenesis, but we assume that the asymmetry was established well before DM freezeout.\footnote{\label{footnote:enhanceAsym}It is possible to enhance the effect of DM assimilation by assuming $\QDirac$ carries a larger asymmetry, which cancels against an \emph{anti}-asymmetry of SM baryons.}

In \Sec{sec:Boltz}, we will be interested in studying the comoving number densities $Y_i \equiv n_i/s$.  It is convenient to package together the scalar modes with the same baryon number
\be
\label{eq:simpScalar2}
Y_{\scalars} \equiv Y_{\QscalarOne} + Y_{\QscalarTwo} \ , \qquad \qquad Y_{\overline{\scalars}} \equiv Y_{\overline{\QscalarOne}} + Y_{\overline{\QscalarTwo}} \ ,
\ee
such that the total baryon asymmetry (before $\scalars$/$\QDirac$ decay) is
\be
\label{eq:etadef}
\etaB \equiv Y_{\scalars}  + Y_{\QDirac}  - Y_{\overline{\scalars}} - Y_{\overline{\QDirac}} \ .
\ee
With this notation, there are no additional factors of two needed to account for the two complex scalars and one Dirac fermion.

As the universe cools, antimatter will be efficiently destroyed, and during DM freezeout, $\etaB \simeq Y_{\scalars}  + Y_{\QDirac}$.  Since $\delta$ in \Eq{eq:deltamass} is positive, $Y_{\scalars} < Y_{\QDirac}$ at all times (i.e.\ the $R$-odd scalars are less abundant than the $R$-even fermions).

\subsection{Late Time Decays}
\label{sec:latedecay}

The heavy $\QDirac$ and $\scalars$ states will decay to the MSSM at late times.  Since the scalars $\scalars$ are $R$-odd, their decays necessarily terminate with the LSP, which in our scenario is the bino.  These decays are responsible for bino DM production, yielding
\be
Y_{\rm DM} \simeq Y_{\scalars} \ ,
\ee
similar to the super-WIMP scenario \cite{Feng:2003xh,Feng:2003uy}.  Corrections to this relation arise because $R$-parity only ensures that $\scalars$ ($\QDirac$) decays will yield an odd (even) number of binos.  Since bino DM requires $Y_{\rm DM} \ll \etaB$, we must ensure that the $R$-even states $\QDirac$ do not produce very many bino pairs, which can happen by arranging the appropriate kinematics.  We also assume that these decays do not violate $U(1)_B$, such that at late times $\etaB$ is the true baryon asymmetry.

\begin{figure}
\begin{center}
\includegraphics[scale=0.55]{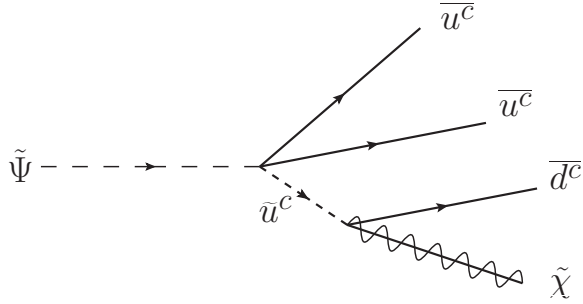} 
\caption{Leading decay of the scalar $\scalars$ to a bino and SM quarks, through a squark.  To avoid problems with the fermion $\QDirac$ decays, we will always take the squark to be off-shell.}
\label{fig:scalardecays}
\end{center}
\end{figure}

With the quantum numbers of $\Q$/$\Qc$, the leading dimension-5 decay operator involving MSSM fields is
\be
\label{eq:decayop}
W_{\rm decay} = \frac{\Q \, \boldsymbol{u^c} \boldsymbol{u^c} \boldsymbol{d^c}}{\Lambda_{\rm decay}} \ .
\ee
Here, $\boldsymbol{u^c}$ and $\boldsymbol{d^c}$ are the electroweak singlet quark multiplets, and we are agnostic as to their flavor structure.\footnote{Since \Eq{eq:decayop} involves an $SU(3)_C$ epsilon tensor, the two $\boldsymbol{u^c}$ fields must have different flavors.}  Since the decay $\scalars \rightarrow \QDirac \bino$ is kinematically forbidden, the dominant decay channel for the scalars $\scalars$ is
\be
\scalars \rightarrow \overline{u^c} \, \overline{u^c} \, \overline{d^c} \, \bino \ ,
\ee
through the Feynman diagram shown in \Fig{fig:scalardecays}.   Note that the $b_M$ mixing term in \Eq{eq:Lsoft} is needed to ensure that both $\QscalarOne$ and $\QscalarTwo$ have allowed decay modes. 

\begin{figure}
\begin{center}
\subfloat[]{\label{fig:fermiondecaysA}\includegraphics[scale=0.43]{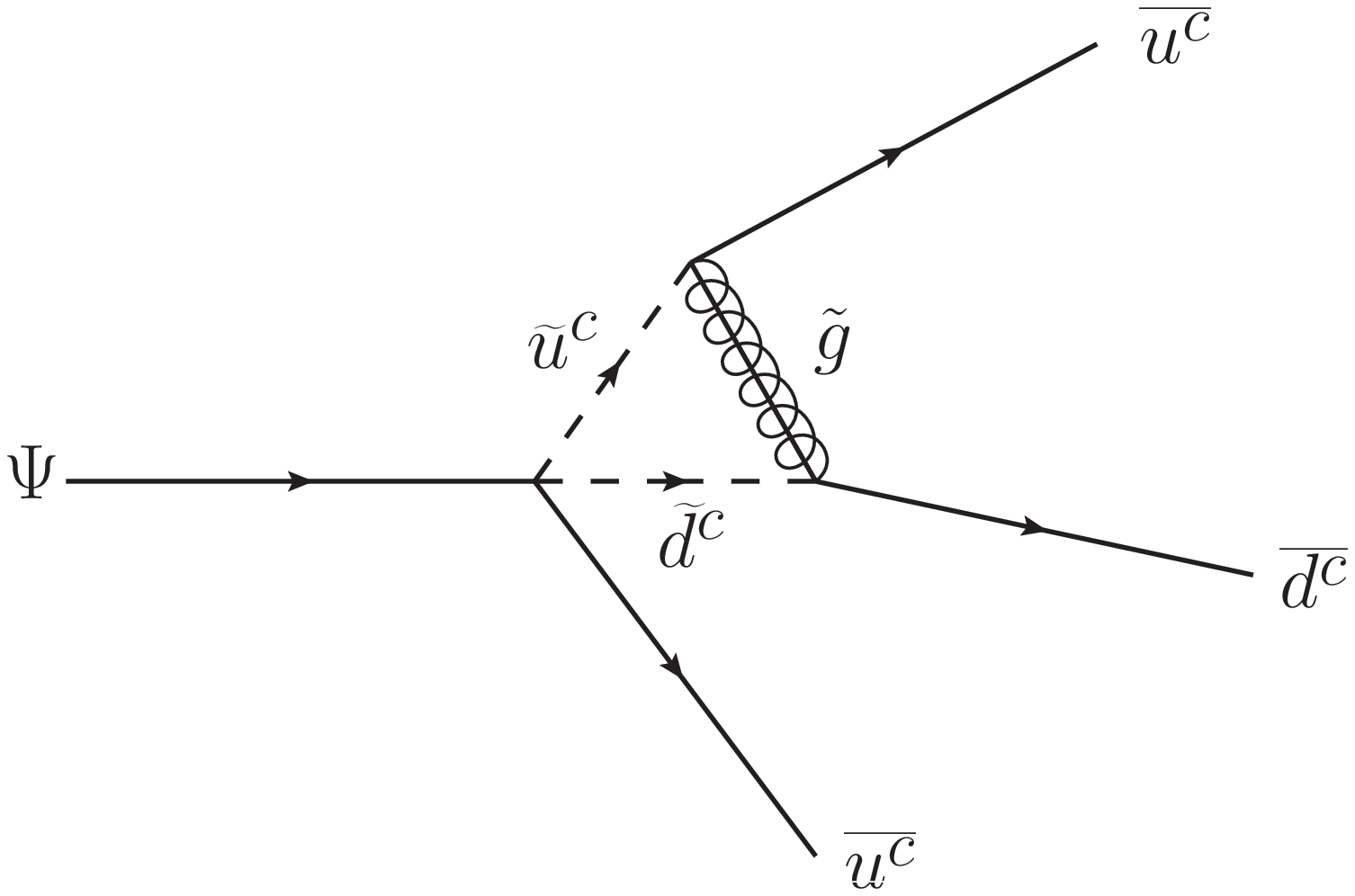}} $\qquad$
\subfloat[]{\label{fig:fermiondecaysB}\includegraphics[scale=0.43]{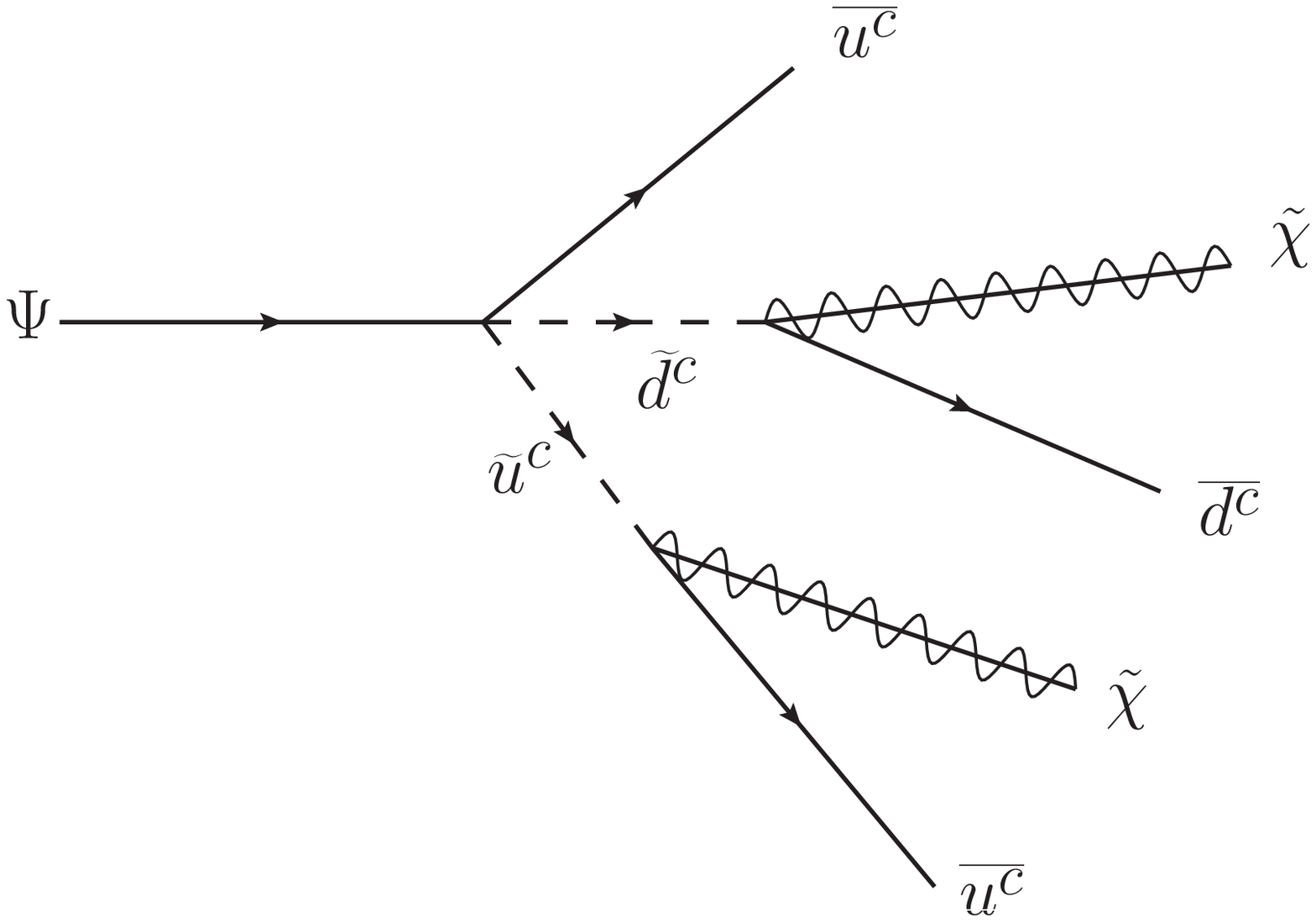}}
\caption{Leading decay channels for the fermion $ \QDirac$:  ({\rm a}) three-body decay to SM quarks through a gluino loop; ({\rm b}) five-body decay to bino pairs and SM quarks through tree-level (off-shell) squarks.  As long as the squarks are kinematically inaccessible, the first process dominates the total width, as needed to forbid bino overproduction through $\QDirac$ decays.}
\label{fig:fermiondecays}
\end{center}
\end{figure}

The leading decays of the fermion $\QDirac$ are
\begin{align}
\QDirac &\rightarrow \overline{u^c} \, \overline{u^c} \, \overline{d^c} \ , \\
\QDirac &\rightarrow \overline{u^c} \, \overline{u^c} \, \overline{d^c} \, \bino 
\bino \ , 
\label{eq:fermiondecays}
\end{align}
through the Feynman diagrams shown in \Fig{fig:fermiondecays}.    If the squarks were kinematically accessible, then decays like $\QDirac \rightarrow \overline{\widetilde{u}^c} \, \overline{\widetilde{u}^c} \, \overline{d^c} \rightarrow \overline{u^c} \,  \overline{u^c} \, \overline{d^c} \, \bino \bino$ would dominate, leading to vast overproduction of binos.  Thus, we must assume the MSSM squarks are not kinematically accessible (or at least phase space suppressed).  That said, even if the squarks are off-shell, the process in \Fig{fig:fermiondecaysB} is still sizable, and we have to make sure that binos originating from this process are a subdominant component of the DM observed today.  From the asymmetry we know that $Y_\QDirac \simeq \etaB \simeq 10^{-10}$, and from observations we know the bino comoving density today is $Y_{ \bino} \simeq 10^{-12}$, therefore the ratio between the partial widths for the two processes in \Fig{fig:fermiondecays} has to be less than $10^{-2}$.  This ratio can be estimated by assuming all the final state particles are massless,\footnote{Since the final state binos are not massless, \Eq{eq:widthratio} is somewhat of an underestimate.} and we assume that the squarks and gluinos have a mass comparable to $M$ for simplicity.  We find
\be
\frac{\Gamma_{\QDirac\rightarrow \overline{u^c} \, \overline{u^c} \, \overline{d^c}}}{\Gamma_{\QDirac \rightarrow \overline{u^c} \, \overline{u^c} \, \overline{d^c} \, \bino \bino}} \simeq \frac{1}{72} \frac{\alpha_Y^2}{\alpha_s^2} \simeq 2 \times 10^{-4} \ ,
\label{eq:widthratio}
\ee
where we have used the dimensionless volume of $n$-body phase space
\be
\Phi^{(n)} = \frac{\Phi^{(n-1)}}{16 \pi^2 \, (n-1) (n-2)} \ , \qquad \Phi^{(2)} = \frac{1}{8 \pi} \ .
\ee
Hence, as long as the squarks cannot be on-shell and the gluino is not too heavy, the loop-suppressed three-body decay $\QDirac \rightarrow  \overline{u^c} \, \overline{u^c} \, \overline{d^c}$ can dominate over the dangerous five-body decay.

In order to have the desired cosmology, $\QDirac$ and $\scalars$ have to be long lived compared to DM freezeout in order to achieve DM assimilation, but short lived compared to Big Bang Nucleosynthesis (BBN) to ensure that the baryon asymmetry is correctly transferred to SM baryons.  Again assuming that the squarks and gluinos have a mass comparable to $M$, we estimate the lifetime of these states as
\begin{align}
(\Gamma_{\scalars})^{-1} &\approx  \left[ \frac{1}{2 M} \frac{\alpha_Y}{\Lambda_{\rm decay}^2} \frac{M^4}{24576 \pi^5}\right]^{-1} \simeq 
7 \times 10^{-3}~\text{sec}~\left(\frac{1~\TeV}{M}\right)^3  \left(\frac{\Lambda_{\rm decay}}{10^8~\TeV}\right)^2  \ ,\\
(\Gamma_{\QDirac})^{-1} &\approx \left[ \frac{1}{2 M} \frac{\alpha_s^2}{(16 \pi^2)^2} \frac{M^2}{\Lambda_{\rm decay}^2} \frac{M^2}{256 \pi^3}\right]^{-1} \simeq 
2 \times 10^{-1}~\text{sec}~\left(\frac{1~\TeV}{M}\right)^3  \left(\frac{\Lambda_{\rm decay}}{10^8~\TeV}\right)^2 \ ,
\end{align}
where the $1/16\pi^2$ factors correspond to loop- or phase-space suppressions.  For the right choice of $\Lambda_{\rm decay}$, these lifetimes are indeed long compared to DM freezeout ($t_{\rm f} \simeq 10^{-8}~\text{sec}$), but short compared to the start of Big Bang Nucleosynthesis ($t_{\rm BBN} \simeq 1~\text{sec}$).  Thus, we can treat $\QDirac$ and  $\scalars$ as stable for the DM freezeout calculation in \Sec{sec:Boltz}, but still have standard BBN.  It might be interesting to study modifications to BBN if $\Lambda_{\rm decay}$ is taken to be larger.

Finally, we note a bound on \Eq{eq:decayop} coming from the fact that this operator can transfer the asymmetry from $\QDirac/\scalars$ to SM baryons via high temperature scattering processes (as opposed to just decays).  This would deplete the asymmetry in $\QDirac/\scalars$ and make DM assimilation less effective.\footnote{See however, footnote \ref{footnote:enhanceAsym}.}  Requiring that the Hubble expansion rate is faster than the scattering rate $\Gamma$ gives a bound on the reheat temperature
\be
T_r <  1.8 \times 10^3 \, \frac{\Lambda_{\rm decay}^2}{M_{{\rm Pl}}} \ .
\ee
Requiring $T_r > M$ to make sure that $\QDirac/\scalars$ have a standard thermal history yields
\be
\Lambda_{\rm decay} > 2.6 \times 10^6~\TeV \left(\frac{M}{1~\TeV}\right)^{1/2} \ ,
\ee
which is consistent with the story above.  A similar constraint would also need to be satisfied in a full theory of $\QDirac$-ogenesis, as well as in the generalizations below.

\subsection{Generalizations}
\label{sec:generalizations}

There are a number of generalization of the model in \Sec{eq:lagrangian} using the same superpotential in \Eq{eq:superpotential}.  As long as $\Q/\Qc$ carries baryon number (not necessarily $\pm1$), has non-zero hypercharge, and has an allowed late-time decay to quarks, then we have a potentially viable model of (pure) bino assimilation.  The $R$-odd states must be heavier than the $R$-even states to avoid overproduction of DM, and the $R$-even state must have a suppressed decay to two binos, perhaps because of kinematics as in \Sec{sec:latedecay}.  A number of possibilities for $\Q/\Qc$ are shown in \Tab{tab:altFields}.  In the cases where $\Q/\Qc$ carry color or electroweak charges, there are additional diagrams compared to \Sec{sec:processes} involving the $SU(3)_C$ and $SU(2)_L$ gauge multiplets.

\begin{table}
\begin{center}
\begin{tabular}{|c|c|c|c|c|}
\hline
$\left(SU(3)_c, SU(2)_L, U(1)_Y\right)$ & $U(1)_B$ & Hierarchy? & Decay Operator \\ \hline\hline
$({\bf 1}, {\bf 1}, +1)$  & $ + 1$ & $m_{\rm scalar} > m_{\rm fermion}$ &$\Q \boldsymbol{u^c} \boldsymbol{u^c} \boldsymbol{d^c}$\\
\hline
$({\bf \overline{3}}, {\bf 1}, +4/3)$  & $ + 2/3$ &$m_{\rm scalar} < m_{\rm fermion}$ &$\Q \boldsymbol{u^c} \boldsymbol{u^c}$\\
$({\bf 3}, {\bf 1}, +2/3)$  & $ + 1/3$ & $m_{\rm scalar} > m_{\rm fermion}$&$\Q \boldsymbol{u^c}$ \\
$({\bf \overline{3}}, {\bf {2}}, -1/6)$  & $ - 1/3$ & $m_{\rm scalar} > m_{\rm fermion}$&$\Q \boldsymbol{q}$ \\
$({\bf 3}, {\bf {3}}, -1/3)$  & $ - 2/3$ &$m_{\rm scalar} < m_{\rm fermion}$ &$\Q \boldsymbol{q} \boldsymbol{q}$\\
$({\bf 1}, {\bf {4}}, -1/2)$  & $ - 1$ & $m_{\rm scalar} > m_{\rm fermion}$ &$\Q \boldsymbol{q} \boldsymbol{q} \boldsymbol{q}$\\
\hline
\end{tabular}
\end{center}
\caption{A non-exhaustive list of alternative quantum numbers for the $\Q$ multiplet in \Sec{eq:lagrangian}.  The first line is the example studied in this paper.  In each case, $\Qc$ has the conjugate quantum numbers to $\Q$.  The mass hierarchy is chosen to make sure that the $R$-odd state is heavier than the $R$-even state.  The decay operators involve the SM quark multiplets, namely the electroweak doublet $\boldsymbol{q}$ and electroweak singlets $\boldsymbol{u^c}$/$\boldsymbol{d^c}$, with unspecified flavor structure.  The bilinear decay operators indicate small mixings with the MSSM states.}
\label{tab:altFields}
\end{table}

Beyond bino DM, there are other examples where singlet DM can be assimilated into a heavier particle in the early universe, and then produced from late time decays. Another SUSY example is a chiral superfield $\boldsymbol{S}$ which is a singlet under the SM gauge group, with the superpotential interaction with heavy fields
\be
W_{S} = \lambda_S\, \boldsymbol{S} \Q \Qc  + m_S  \boldsymbol{S}^2 \ .
\ee
The $R$-odd fermion singlino $\widetilde{s}$ DM would be assimilated into the $R$-odd scalars $\scalars$.    A key ingredient is an annihilation channel for the $R$-odd states that conserves baryon number
\be
\scalars \scalars \rightarrow \QDirac \QDirac \ ,
\ee
which can occur via $t$-channel gaugino and singlino exchange.  The late-time decays which populate DM and baryons would occur through operators like \Eq{eq:decayop}.

Finally, while the new states need to carry some kind of asymmetry for assimilation to be effective, it need not be the baryon asymmetry.  Since the new states eventually need to decay, this asymmetry would persist until today, so the baryon asymmetry is the natural choice given phenomenological constraints.\footnote{One amusing possibility would be if the new states could carry a DM asymmetry (as in asymmetric DM \cite{Kaplan:2009ag}).  Assimilation could then be used to help eliminate the symmetric component.}

\section{Bino Assimilation in the Early Universe}
\label{sec:Boltz}

In this section, we study in detail how bino assimilation happens in the early universe.   After describing the relevant processes, we derive the Boltzmann equations for the species evolution, and present semi-analytical and numerical results.  Various details appear in the appendices.

\subsection{Relevant Processes}
\label{sec:processes}

We start the thermal history from early times, when the temperature of the universe is higher than the superpotential mass $M$ but the baryon asymmetry is already stored in $\Q$/$\Qc$.  At this stage, the asymmetry in \Eq{eq:etadef} is largely irrelevant, and particles and antiparticles democratically populate the universe.  Once the temperature drops below the mass $M$, the antiparticles  $\overline{\QDirac}$ and $\overline{\scalars}$ quickly disappear as a consequence of annihilation, as shown in \App{app:LWasymm}.  The antiparticles can be neglected at lower temperatures, leaving only $\QDirac$ and  $\scalars$ carrying the baryon asymmetry:
\be
\etaB = Y_{\scalars}  + Y_{\QDirac} \ .
\ee

\begin{figure}
\begin{center}
\subfloat[]{\includegraphics[scale=0.5]{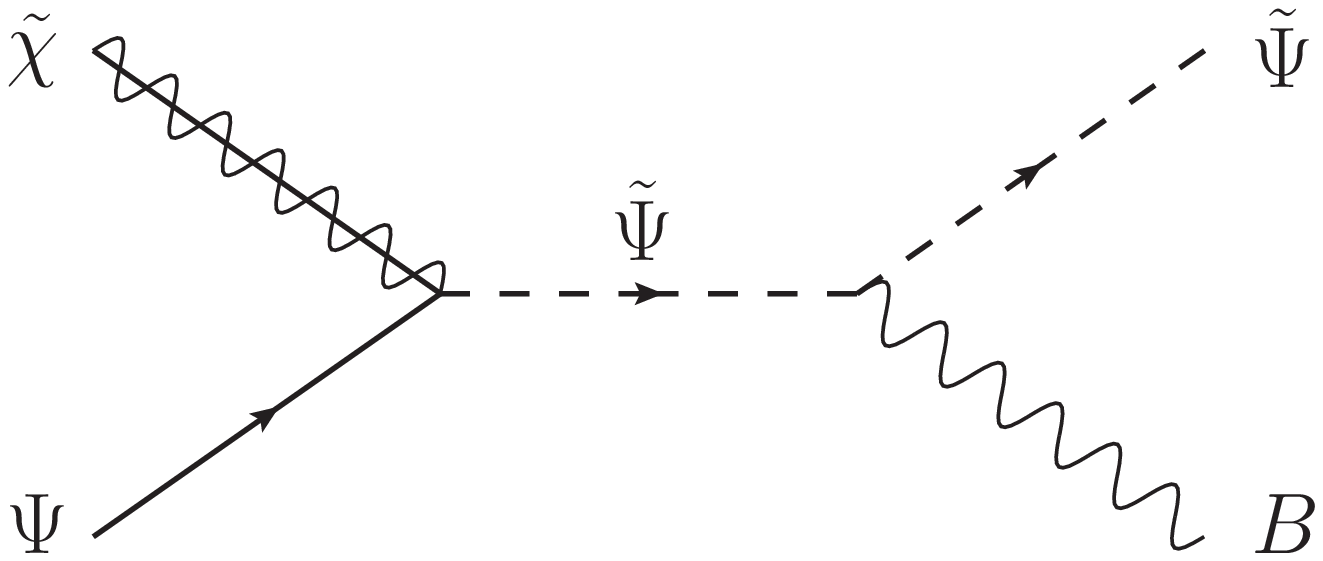}} $\qquad$
\subfloat[]{\includegraphics[scale=0.5]{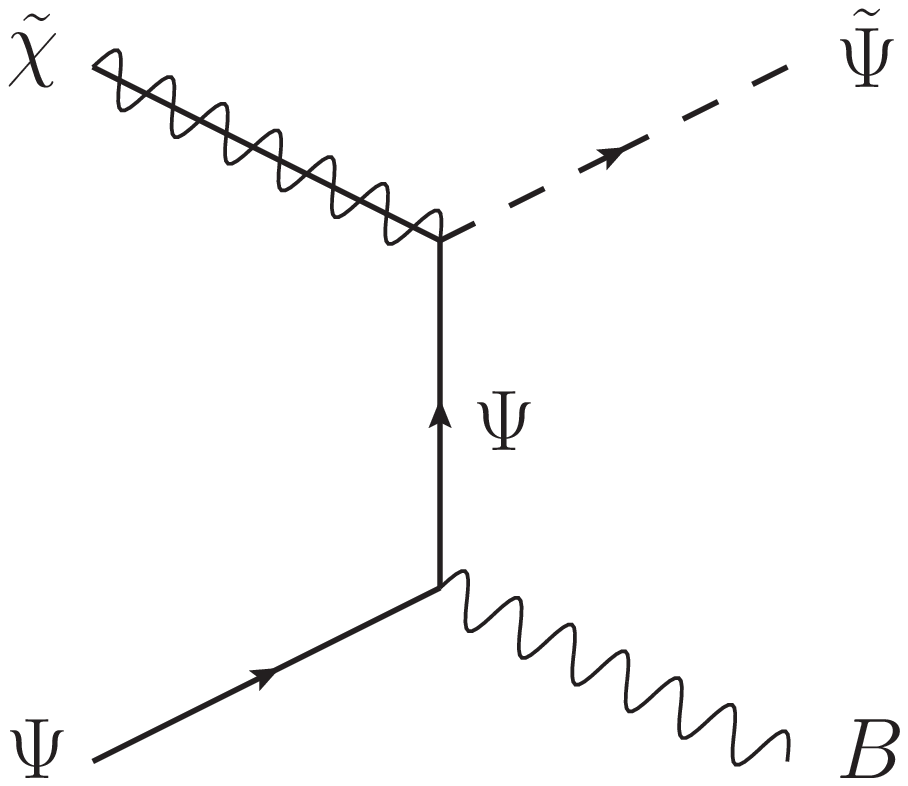}}
\end{center}
\caption{Assimilation of the bino $\bino$, converting fermions $\QDirac$ to scalars $\scalars$.  The diagrams are ({\rm a}) $s$-channel scalar exchange; and ({\rm b}) $t$-channel fermion exchange. The field $B$ denotes the hypercharge gauge boson.}
\label{fig:Feynassimilation}
\end{figure}

Given the absence of antiparticles, there are only three classes of processes relevant for DM freezeout. The first relevant process is assimilation shown in \Fig{fig:Feynassimilation},
\be
\label{eq:process1}
\text{Assimilation}:  \qquad  \bino  \, \QDirac \; \rightarrow \; \scalars \, B \ ,
\ee
where $B$ is the hypercharge gauge boson.\footnote{Though $B$ is a linear combination of the mass eigenstates $\gamma$ and $Z$, we ignore the $Z$ mass for simplicity.}  Since the $\QDirac$ and $\scalars$ states decay much later than DM freezeout and they satisfy the triangle inequality $m_{\scalars} < m_{\QDirac} + m_{\bino}$, this is an example of a semi-annihilation reaction \cite{D'Eramo:2010ep}.  In this process, the bino $\bino$ gets destroyed, but its odd $R$-parity is stored in the final state $\scalars$.  In other words, DM has been assimilated.

\begin{figure}
\begin{center}
\subfloat[]{\includegraphics[scale=0.5]{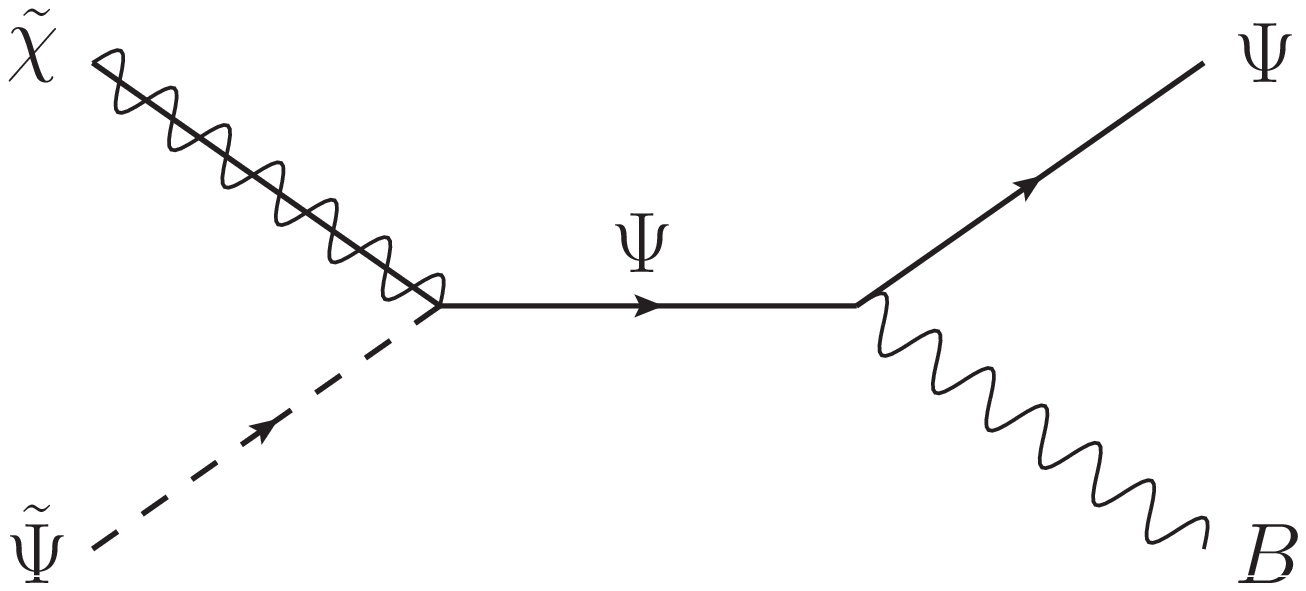}} $\qquad$
\subfloat[]{\includegraphics[scale=0.5]{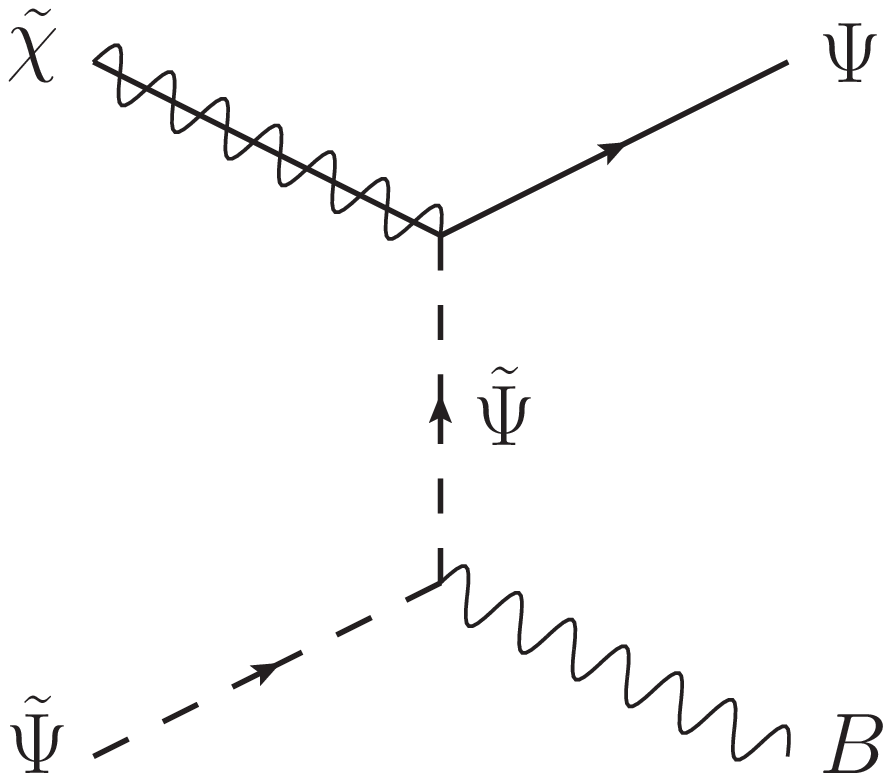}}
\end{center}
\caption{Destruction of $R$-odd particles for: ({\rm a}) $s$-channel fermion exchange; and ({\rm b}) $t$-channel scalar exchange.}
\label{fig:FeyndestrRodd}
\end{figure}

The second relevant process is destruction of $R$-odd particles shown in \Fig{fig:FeyndestrRodd},
\be
\label{eq:process2}
\text{Destruction}:  \qquad \bino  \, \scalars  \; \rightarrow \;  \QDirac \, B \ .
\ee
This is also a semi-annihilation process, and the associated diagrams can be obtained by crossing symmetry from \Fig{fig:Feynassimilation}.  The common features of these two process, shared with all semi-annihilation reactions, is that they are never phase space suppressed.  Thus, they are very effective in the early universe at keeping the bino in thermal equilibrium, causing near total bino depletion. 

\begin{figure}
\begin{center}
\includegraphics[scale=0.5]{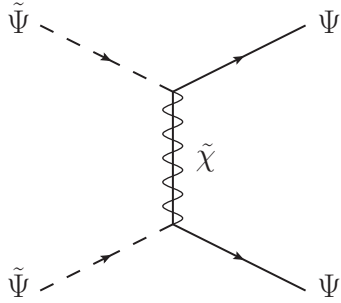} 
\caption{Conversion of $R$-odd particles to $R$-even particles through $t$-channel bino exchange.}
\label{fig:speconv}
\end{center}
\end{figure}

While \Eq{eq:process2} is effective at keeping the $\bino$ in thermal equilibrium, it is not the most effective way to eliminate $R$-odd particles, since both assimilation and destruction are controlled by the same bino abundance.  The third and last relevant process is $R$-parity conversion shown in \Fig{fig:speconv},
\be
\label{eq:process3}
\text{Conversion}: \quad \scalars \, \scalars \; \rightarrow \; \QDirac \, \QDirac \ .
\ee
As long as $\scalars$ are heavier than $\QDirac$, this is most effective way to get rid of $R$-odd particles in the early universe, despite the fact that this process is phase-space suppressed.  In the semi-analytical study in \Sec{sec:semianalytic}, this conversion process will be solely responsible for the freezeout of the scalars. Finally, when the scalars and fermions decay at late times, the binos are repopulated with density $Y_{\scalars}$.

\subsection{Boltzmann Equations}

The Boltzmann equations describe the evolution of the comoving number densities of the bino $\bino$ and the scalars $\scalars$.  During DM freezeout, the comoving number density of fermions $\QDirac$ is given by baryon number conservation
\be
\label{eq:latetimeEtaBrelation}
Y_{\QDirac} = \etaB - Y_{\scalars} \ ,
\ee
so we need not have a separate Boltzmann equation for $Y_{\QDirac}$.

We introduce a dimensionless time variable $x$ and cross sections $\lambda_i$,
\be
\label{eq:rescaleDefs}
x = \frac{M}{T} \ , \qquad \lambda_i = \frac{s(T = M)}{H(T = M)} \langle \sigma v_{{\rm rel}} \rangle_i  \ ,
\ee
where $T$ is temperature, $H$ is the Hubble parameter, and $s$ is the entropy density of relativistic degrees of freedom.  The equilibrium comoving density for the bino is
\be
\Yeq_{\bino} = \frac{n^{{\rm eq}}_{\bino}}{s} = \frac{g_{\bino}}{g_{* s}} \frac{45}{4 \pi^4} \left(\frac{m_{\bino}}{M}\right)^2x^2 \, K_2\left[\frac{m_{\bino}}{M} x \right] \ ,
\ee
where $g_{\bino} = 2$ and $K_2[x]$ is the modified Bessel function.  To express the equilibrium distribution for the new degrees of freedom $\QDirac$ and $\scalars$, it is convenient to introduce the ratio between the equilibrium distributions which is not affected by the asymmetry
\be
r^{\rm eq}_\Psi(x) \equiv \frac{Y^{{\rm eq}}_{\scalars} (x)}{Y^{{\rm eq}}_{\QDirac}(x)}  = \left(1 + \frac{\delta}{M}\right)^{3/2} e^{-(\delta/M)x}  \ ,
\ee
using the parameter $\delta$ from \Eq{eq:deltamass}.  The equilibrium distributions are
\be
\Yeq_{\scalars}(x) = \frac{r^{\rm eq}_\Psi(x)}{1 +  r^{\rm eq}_\Psi(x)} \, \etaB \ , \qquad \Yeq_{\QDirac} (x) = \frac{1}{1 +  r^{\rm eq}_\Psi(x)} \, \etaB \ .
\label{eq:Ypsieq}
\ee

The Boltzmann equations, derived in detail in \App{app:Boltzmann}, are 
\be
\begin{split}
& \frac{d Y_{\bino}}{d x} = 
- \frac{\lambda_{\rm assim}}{x^2} \left[Y_{\bino} Y_{\QDirac} - \frac{\Yeq_{\bino} \Yeq_{\QDirac}}{\Yeq_{\scalars} } Y_{\scalars}  \right]
- \frac{\lambda_{\rm dest}}{x^2}  \left[Y_{\bino} Y_{\scalars} - \frac{\Yeq_{\bino} \Yeq_{\scalars}}{\Yeq_{\QDirac}} Y_{\QDirac} \right]  \ , 
\end{split}
\label{eq:Boltz3eqsA}
\ee
\be
\begin{split}
\frac{d Y_{\scalars}}{d x} = & \,
- \frac{\lambda_{\rm assim}}{x^2} \left[\frac{\Yeq_{\bino} \Yeq_{\QDirac}}{\Yeq_{\scalars} } Y_{\scalars} - Y_{\bino} Y_{\QDirac}  \right]  
- \frac{\lambda_{\rm dest}}{x^2}  \left[Y_{\bino} Y_{\scalars} - \frac{\Yeq_{\bino} \Yeq_{\scalars}}{\Yeq_{\QDirac}} Y_{\QDirac} \right] 
 \\ & \, -  \frac{\lambda_{\rm conv}}{x^2} \left[Y^2_{\scalars} - \left(\frac{\Yeq_{\scalars}}{\Yeq_{\QDirac}}\right)^2 Y^2_{\QDirac} \right] \ .
\end{split}
\label{eq:Boltz3eqsB}
\ee
The effective cross section parameters can be expressed in terms of dimensionless cross sections $\lambda_i$ for the fundamental processes in our model
\be
\begin{split}
\lambda_{\rm assim}  & \equiv \lambda_{\bino \QDirac \; \rightarrow \; \scalars B} = 2 \; \lambda_{\bino \QDirac \; \rightarrow \; \Qscalar B}  \ , \\ 
\lambda_{\rm dest} & \equiv \lambda_{\bino \scalars \; \rightarrow \; \QDirac B} =  \lambda_{\bino \, \Qscalar \; \rightarrow \; \QDirac B} \ , \\
\lambda_{\rm conv}   & \equiv \lambda_{\scalars \scalars \; \rightarrow \; \QDirac\QDirac} = \frac{\lambda_{\Qscalar \Qscalar \; \rightarrow \; \QDirac\QDirac} + \lambda_{\Qscalar \overline{\Qcscalar} \; \rightarrow \; \QDirac\QDirac}}{2} \ .
\end{split}
\label{eq:lambdas}
\ee
Here, we are relying on the simplified scalar notation in \Eqs{eq:simpScalarNotation}{eq:simpScalar2}, and the resulting factors of two are tracked in detail in  \App{app:Boltzmann}.

\subsection{Semi-Analytical Solution}
\label{sec:semianalytic}

In general, the two coupled Boltzmann equations in \Eqs{eq:Boltz3eqsA}{eq:Boltz3eqsB} have no closed form solution, and we will solve them numerically in the next subsection.  However, there is a particular region of parameter space where the number density evolution can be understood semi-analytically.   This occurs if the assimilation and destruction processes are effective enough to keep $\bino$ in thermal equilibrium, such that $\QDirac$ and $\scalars$ freezeout when there are very few $\bino$ around.  In this case, we can focus on the $\QDirac$ and $\scalars$ system alone, and the bino DM density today is determined directly from the freezeout density of the scalars $\scalars$.

If we can neglect the bino number density $Y_{\bino}$, the assimilation and destruction processes are subdominant, and the number density of scalars $\scalars$ is approximately described by the conversion process alone
\be
\frac{d Y_{\scalars}}{d x} = - \frac{\lambda_{\rm conv}}{x^2} \left[Y^2_{\scalars} - (r^{\rm eq}_\Psi)^2 \left(\etaB - Y_{\scalars} \right)^2 \right] \ .
\ee
The above equation can be solved semi-analytically, in analogy with the Lee-Weinberg calculation \cite{Lee:1977ua,Kolb:1990vq}. At early times, the conversion reactions keep the $\scalars$ particles in thermal equilibrium. Eventually, a freezeout point is reached where the reaction rate $\Gamma_{\rm conv} = n_{\scalars}^{\rm eq} \langle \sigma v_{{\rm rel}} \rangle_{\rm conv}$ cannot compete with the Hubble expansion rate $H$.
We define the freezeout point $x_f$ by
\be
\label{eq:xfdef}  
 H(x_f) \equiv c \, \Gamma_{\rm conv}(x_f) \ ,
 \ee
 where $c$ is a constant of order $1$.  This yields the relation $c \, Y^{{\rm eq}}_{\scalars} \,\lambda_{\rm conv} = x_f$,
 or equivalently 
\be
x_f = \frac{M}{\delta} \log\left[\frac{c \, \eta_B \, \lambda_{\rm conv}}{x_f}  -1 \right] + \frac{3}{2} \frac{M}{\delta}\log\left[1 + \frac{\delta}{M}\right] \ .
\label{eq:xf}  
\ee
After freezeout, the $r^{\rm eq}_\Psi$ factors in the Boltzmann equation are exponentially suppressed, and the evolution is straightforward.  Ignoring the freezeout boundary condition, the asymptotic value for $Y_{\scalars}$ is
\be
Y_{\scalars}(\infty) = \frac{x_f}{\lambda_{\rm conv}} \ .
\ee
 
Since the scalars $\scalars$ will eventually decay to the bino $\bino$, and since (almost) all the DM density comes from this decay, the mass density of DM today is
\be
\rho_{\bino, 0} = m_{\bino} \, s_0 \, Y_{\scalars}(\infty) \ .
\ee
As usual we can express the DM density as a function of the critical energy density
\be
\Omega_{\bino} h^2 = \frac{1.07 \times 10^9~\GeV^{-1}}{\sqrt{g_*} \, M_{{\rm Pl}}}  \, \frac{x_f}{\langle \sigma v_{{\rm rel}} \rangle_{\rm conv}} \, \frac{m_{\bino}}{M} \ ,
\label{eq:Omegasemi}
\ee
where the effective cross section $\langle \sigma v_{{\rm rel}} \rangle_{\rm conv}$ is related to the parameter $\lambda_{\rm conv}$ by \Eqs{eq:rescaleDefs}{eq:lambdas}.  This is the final semi-analytical result.  The important difference between \Eq{eq:Omegasemi} and the familiar Lee-Weinberg result \cite{Lee:1977ua,Kolb:1990vq} is the ratio $m_{\bino}/M$.  This arises because the particle freezing out is not the same as the one comprising DM today, analogous to the super-WIMP case \cite{Feng:2003xh,Feng:2003uy}.

To get a sense for the range of parameters that lead to the desired DM density, we compute the conversion cross section for our particular model in \App{app:amplitudes}.  In the $\delta \ll M$ limit
\be
\langle \sigma v_{{\rm rel}} \rangle_{\rm conv}  \simeq \frac{2 \pi \alpha_Y^2}{m_{\bino}^2} \, \frac{m_{\rm soft}}{M} \ ,
\label{eq:crosec}
\ee
where $m_{\rm soft}$ is related to $\delta$ in \Eq{eq:deltamass}.  Combining \Eqs{eq:Omegasemi}{eq:crosec}, we find
\be
\Omega_{\bino} h^2 \simeq  0.1 \, \left(\frac{m_{\bino}}{500~\GeV}\right)^3 \,  \left(\frac{m_{\rm soft}}{200~\GeV}\right)^{-1} \, \frac{x_f}{80} \ ,
\label{eq:Omegaourmodel}
\ee
where we have used $g_* = 106.75$ and $M_{{\rm Pl}} = 1.22 \times 10^{19}~\GeV$.  This is only an approximate result, valid in the $\delta \ll M$ limit and assuming that $Y_{\bino}$ can be neglected.  For reasonable values of $m_{\bino}$ and $m_{\rm soft}$, we obtain the desired relic density.  The dependence on $M$ comes in through $x_f$ in \Eq{eq:xf}.

\subsection{Numerical Results}
\label{sec:results}

\begin{table}
\begin{center}
\begin{tabular}{|c||c|c|c|}
\hline
Benchmark & $M$ & $m_{\bino}$& $m_{\rm soft}$ \\
\hline \hline
A&  $1.4~\TeV$ &  $180~\GeV$ &   $200~\GeV$\\ 
B&  $500~\GeV$ & $390~\GeV$ &  $200~\GeV$\\ 
C&  $900~\GeV$ &  $180~\GeV$& $350~\GeV$\\ 
D&  $600~\GeV$ &  $390~\GeV$ & $350~\GeV$\\
\hline
\end{tabular}
\end{center}
\caption{Four benchmark points that yield the observed DM relic abundance $\Omega_{\rm DM} h^2 \simeq 0.1$.}
\label{tab:benchmarks}
\end{table}

\begin{figure}
\begin{center}
\subfloat[]{\label{fig:BoltznumA} \includegraphics[scale=0.55]{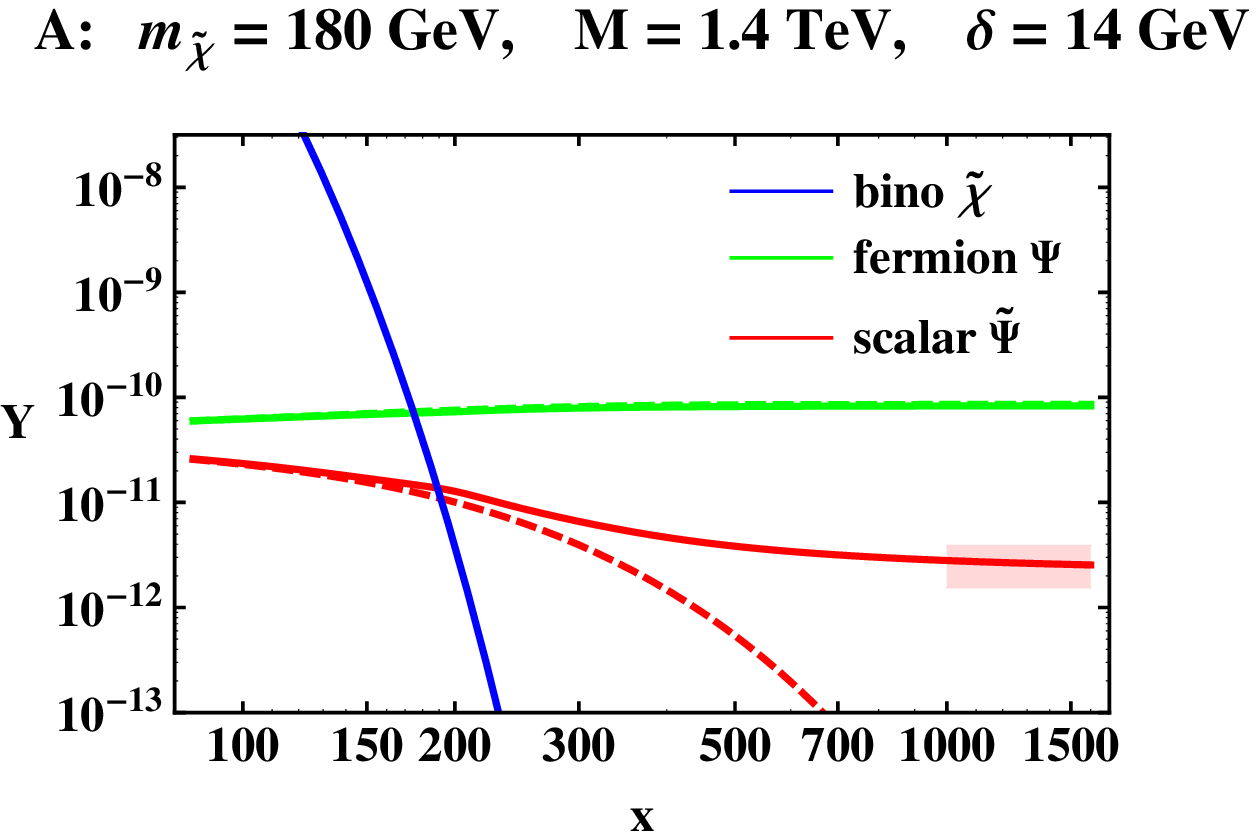}} $\qquad$
\subfloat[]{\label{fig:BoltznumB} \includegraphics[scale=0.55]{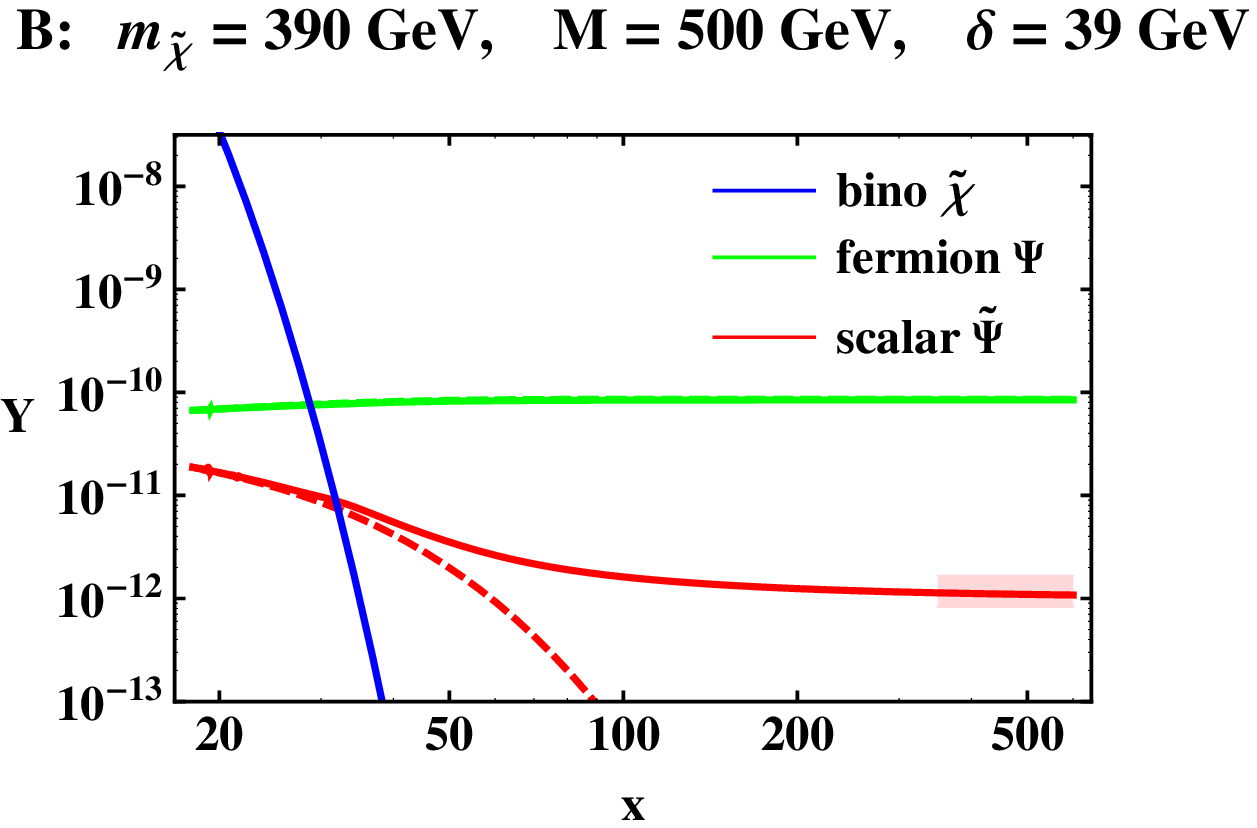}}  \\
\subfloat[]{\label{fig:BoltznumC} \includegraphics[scale=0.55]{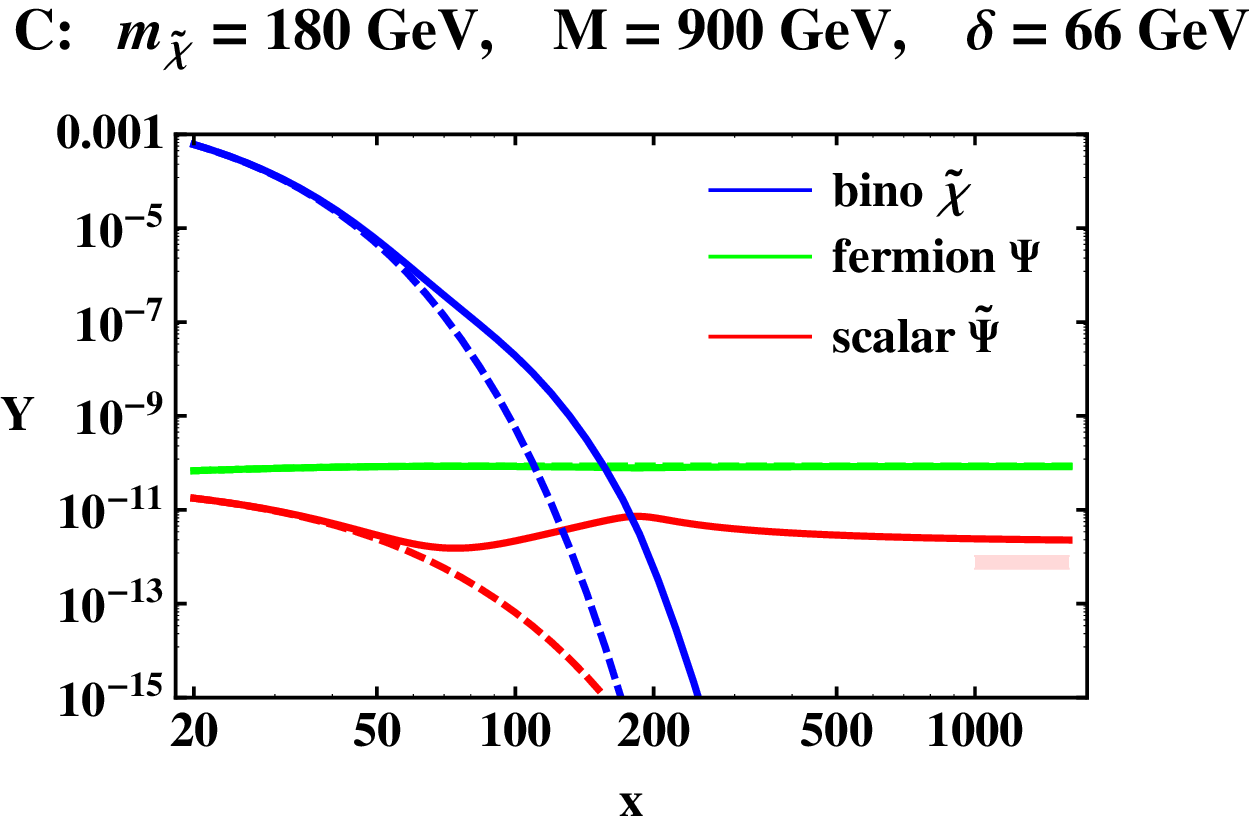}} $\qquad$
\subfloat[]{\label{fig:BoltznumD} \includegraphics[scale=0.55]{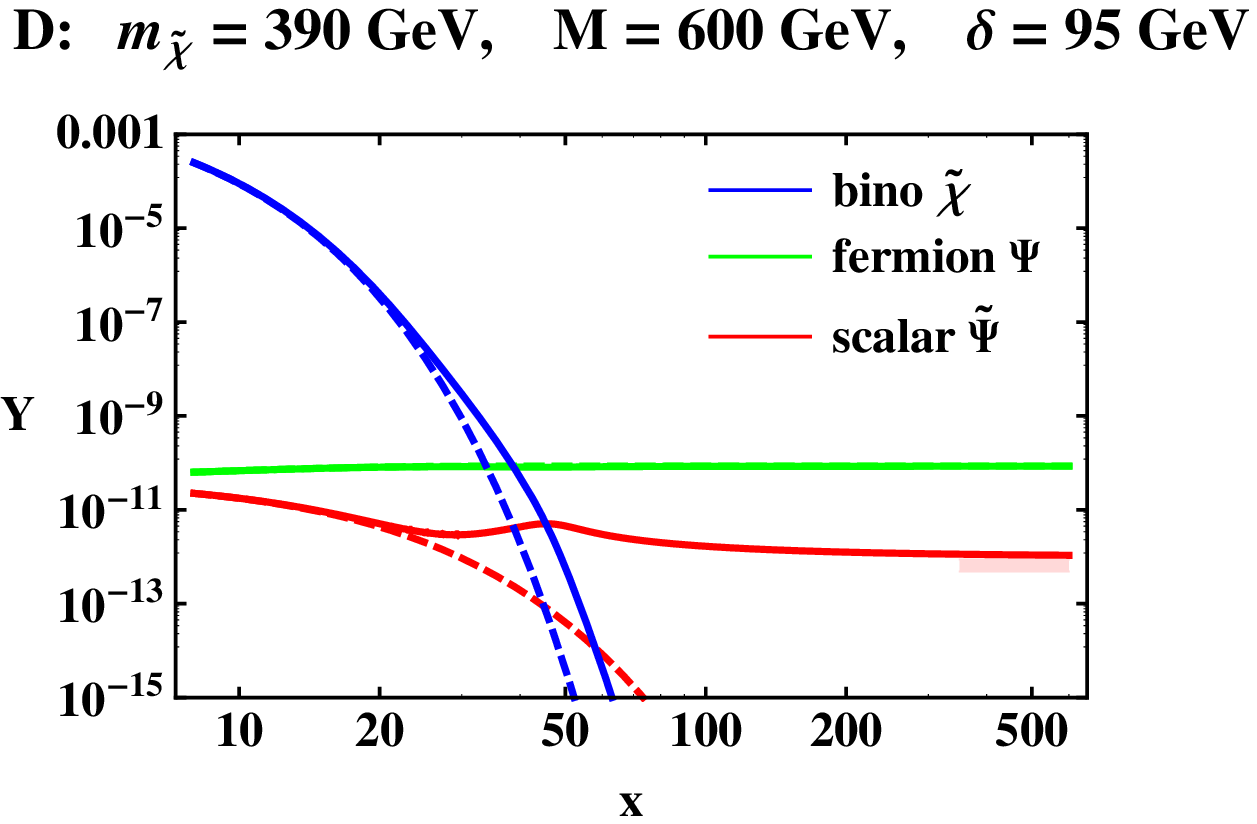}}
\end{center}
\caption{Numerical results for the Boltzmann system in the four benchmark scenarios from \Tab{tab:benchmarks}.  The top row has $m_{\rm soft} \simeq 200~\GeV$ while the bottom row has $m_{\rm soft} \simeq 350~\GeV$.  The left column has $m_{\bino} \ll M$ while the right column has $m_{\bino} \simeq M$.  In all cases the parameters are chosen to reproduce the observed value of the DM density, $\Omega_{\rm DM} h^2 \simeq 0.1$. We plot the comoving number densities for the bino $\bino$ (blue), Dirac fermion $\QDirac$ (green) and scalars $\scalars$ (red).  We also plot the equilibrium distributions, using the same colors but dashed lines.  The light red band shows the semi-analytical solution for the asymptotic scalars comoving density from \Eq{eq:Omegaourmodel}, sweeping the range $c = [0.1, 2]$.  As discussed in the text, for benchmarks C and D, the semi-analytical result is not valid since the bino density is large at freezeout. At late times, the bino number density is given by $Y_{\scalars}$.}
\label{fig:Boltznum}
\end{figure}

We now solve the Boltzmann system in \Eqs{eq:Boltz3eqsA}{eq:Boltz3eqsB} numerically, using the thermally averaged cross section in \App{app:amplitudes}.  We will first study four representative benchmark points given in \Tab{tab:benchmarks}.  In all four cases, we have selected values for the scalar soft mass $m_{\rm soft}$, the superpotential mass $M$, and the bino mass $m_{\bino}$ that reproduce the observed relic density, $\Omega_{\rm DM} h^2 \simeq 0.1$.  For benchmarks A and C, the bino is much lighter than $\QDirac$/$\scalars$, and for benchmarks B and D, the spectrum is more compressed.  In \Fig{fig:Boltznum}, we show the evolution of the comoving number densities.  

Benchmarks A and B have a common value $m_{\rm soft} \simeq 200~\GeV$, and their evolution is shown in \Figs{fig:BoltznumA}{fig:BoltznumB}.   For these two benchmarks, the bino distribution follows very closely its equilibrium value, which implies the semi-analytical analysis in \Sec{sec:semianalytic} will be valid.  The bino is kept in thermal equilibrium due to the very effective assimilation/destruction reactions
\be
 \bino \QDirac \; \rightarrow \; \scalars B \ , \qquad \bino \scalars \; \rightarrow \; \QDirac B \ .
\ee
The above processes almost completely wash out the binos, which are ``assimilated'' into the scalars $\scalars$.   Turning to the heavy states, if thermal equilibrium were maintained between $\QDirac$ and $\scalars$, all the scalars $\scalars$ would eventually be suppressed, leaving no $R$-odd particles around today, and thus no DM.  However, the expansion of the universe prevents that from happening, and the scalar comoving density freezes out. From the semi-analytical result in \Eq{eq:xf}, we know that the freezeout point scales as $x_f \simeq M / \delta$, with a logarithmic dependence on the other variables, and this dependence of $x_f$ can be seen by comparing the departure from equilibrium between \Fig{fig:BoltznumA} and \Fig{fig:BoltznumB}.

We next consider benchmarks C and D in \Figs{fig:BoltznumC}{fig:BoltznumD}, which have a larger value $m_{\rm soft} \simeq 350~\GeV$.  We immediately notice a substantial difference from the previous case, such that the semi-analytical solution cannot be trusted.  Increasing the soft mass yields a dramatic decrease in the $\scalars$ thermal equilibrium distribution from \Eq{eq:Ypsieq}, because of the larger mass splitting appearing in the exponential.\footnote{The cross sections $\lambda_i$ appearing in the Boltzmann system in \Eqs{eq:Boltz3eqsA}{eq:Boltz3eqsB} are not appreciably affected by the higher soft mass value.}  As a result, the overall interaction rates are smaller than in the previous case, and there is an early departure from thermal equilibrium (``freezeout'') while the $\bino$ density is orders of magnitude above the $\QDirac$/$\scalars$ density.  The subsequent evolution of the scalar particles $\scalars$ involves two competing effects.   Immediately after freezeout, the assimilation process starts to regenerate $\scalars$, since there is a large density of $\bino$ that drives the assimilation reaction.  Eventually, the $\bino$ density decreases sufficiently that the destruction and conversion processes can reduce the number of $\scalars$ until the asymptotic value is reached.  While the DM dynamics is richer and less straightforward for large values of the $m_{\rm soft}$, the overall picture is left unchanged:  the scalars $\scalars$ assimilate the bino $\bino$, and late-time scalar decays produce the DM observed today.

\begin{figure}
\begin{center}
\subfloat[]{\label{fig:BoltzunifA}\includegraphics[scale=0.57]{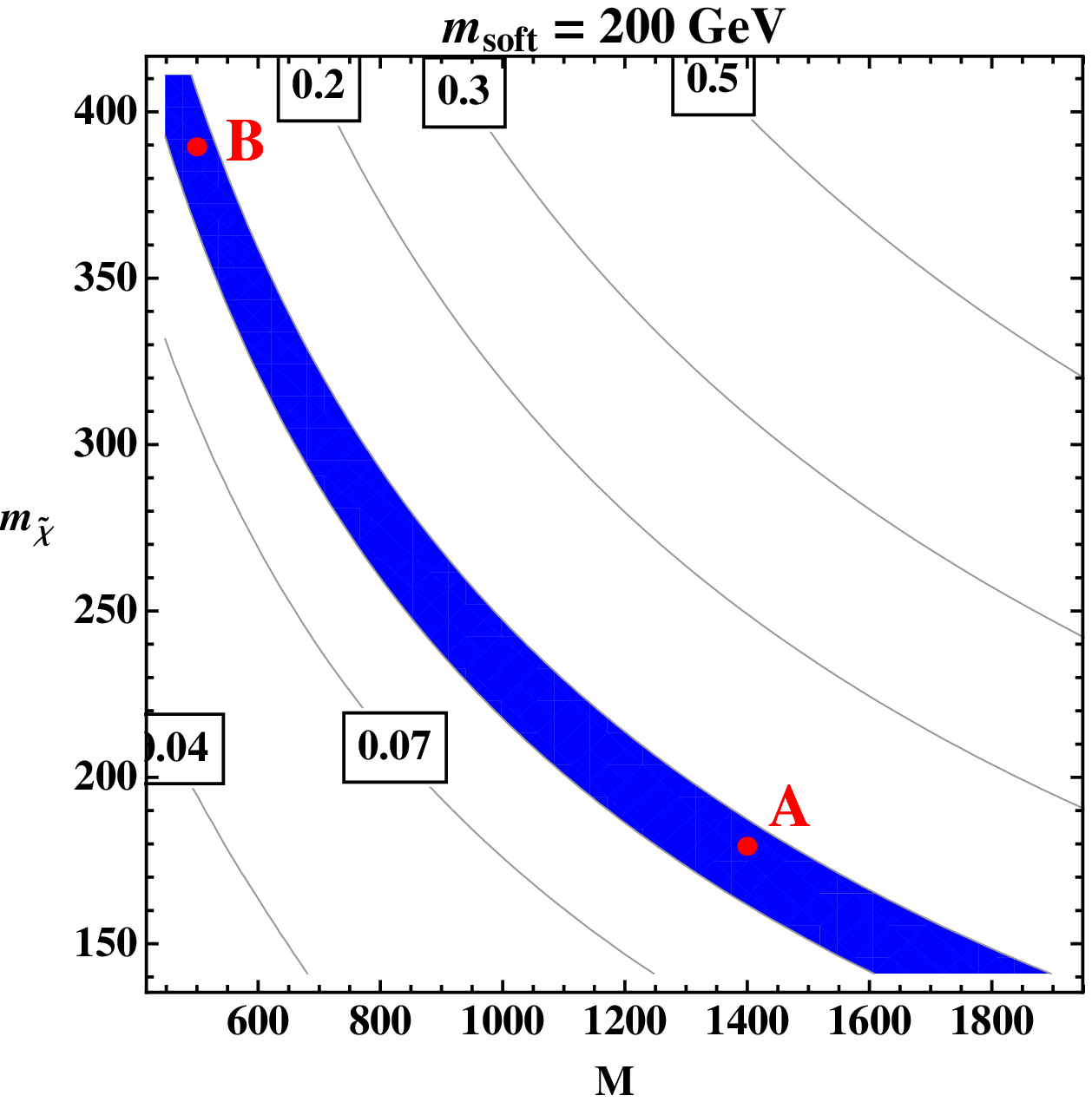}} $\qquad$
\subfloat[]{\label{fig:BoltzunifB}\includegraphics[scale=0.57]{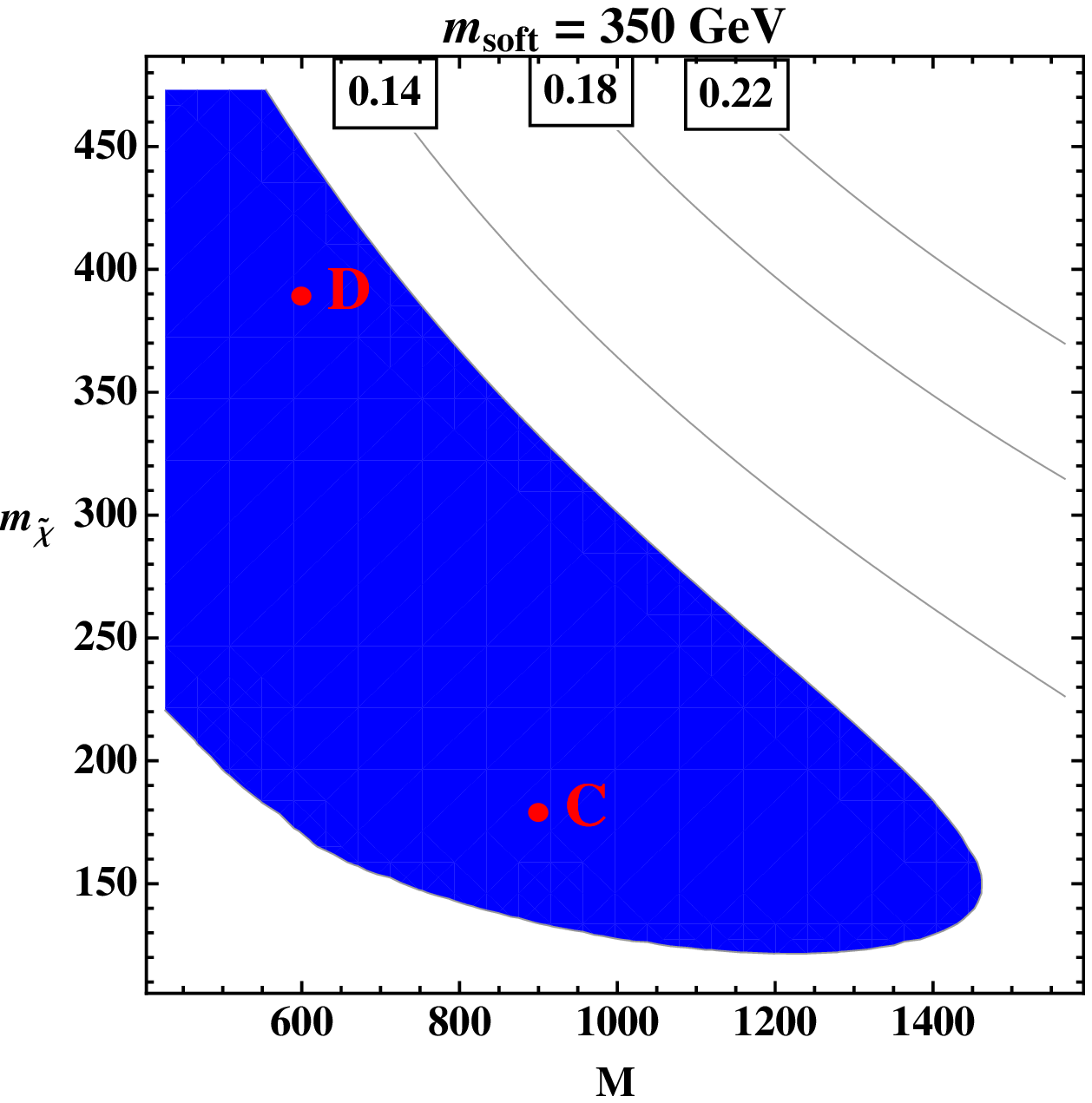}} \\ ~\\ ~\\
\subfloat[]{\label{fig:BoltzunifAbis}\includegraphics[scale=0.57]{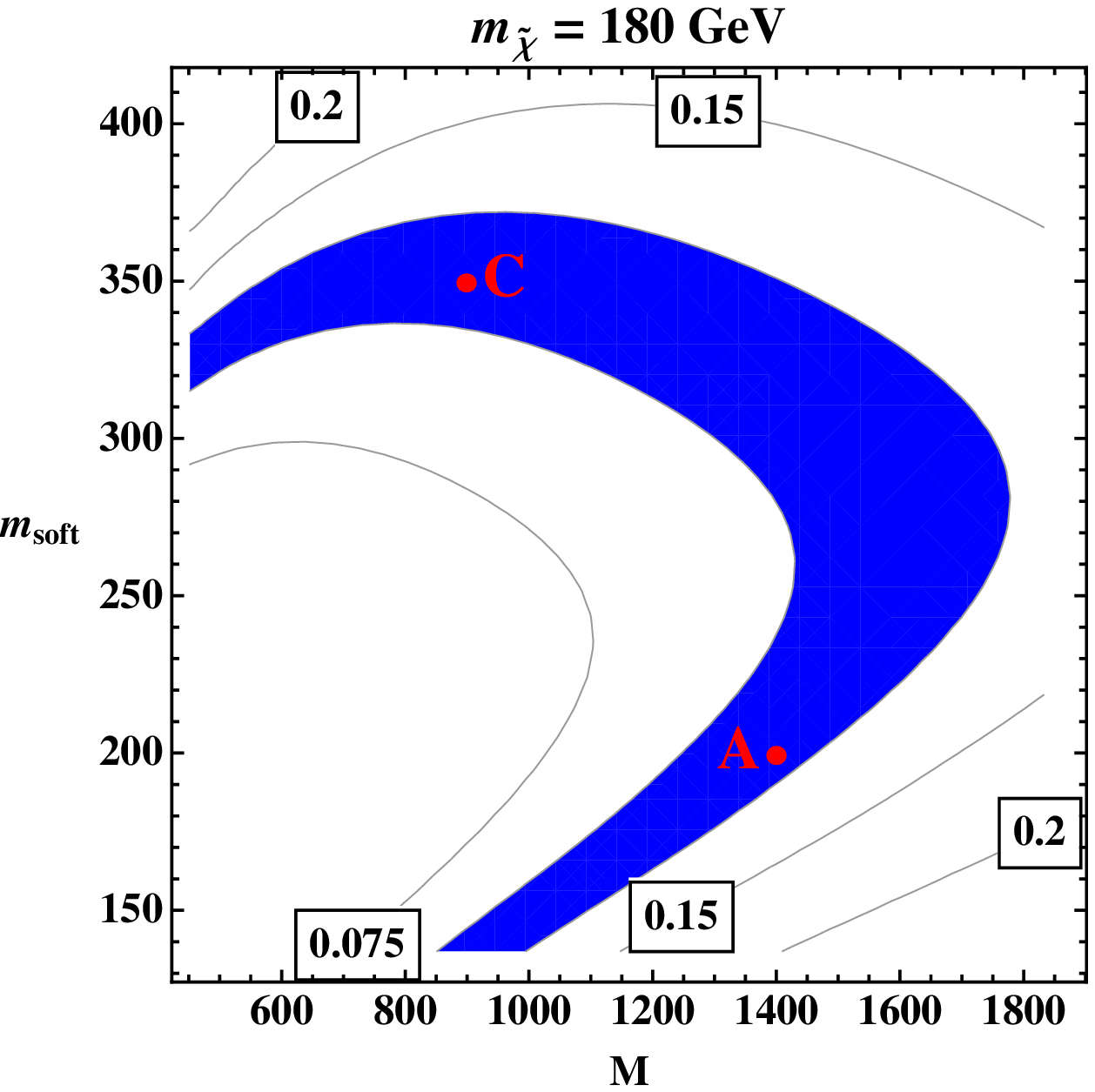}} $\qquad$
\subfloat[]{\label{fig:BoltzunifBbis}\includegraphics[scale=0.57]{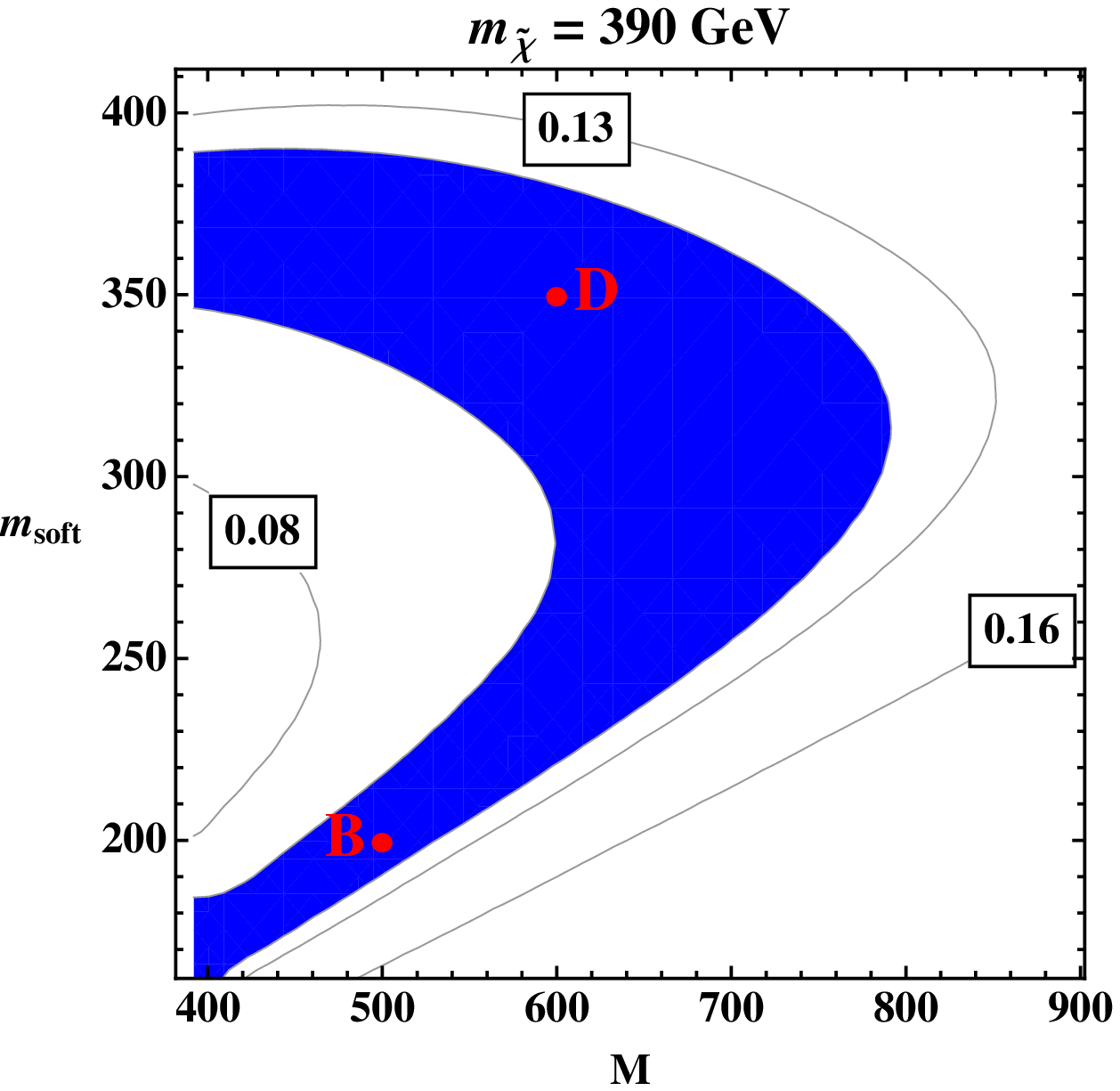}}
\end{center}
\caption{DM relic density, considering slices through the $\left(M, m_{\bino}, m_{\rm soft}\right)$ parameter space.  The shaded blue regions identify the WMAP $95\%$ CL region, $0.0975 \leq \Omega_{{\rm DM}} h^2 \leq 0.1223$, and the contour lines correspond to constant DM density. The red dots indicate the benchmark points in \Tab{tab:benchmarks}.  Top row:  relic density in the $\left(M, m_{\bino}\right)$ plane for: (a) $m_{\rm soft} = 200~\GeV$; (b) $m_{\rm soft} = 350~\GeV$.  Bottom row:  relic density in the $\left(M, m_{\rm soft}\right)$ plane for: (a) $m_{\bino} = 180~\GeV$; (b) $m_{\bino} = 390~\GeV$.}
\label{fig:Boltzunif}
\end{figure}

Once the quantum numbers of $\Q$/$\Qc$ are fixed, the DM density depends on only three parameters: the superpotential mass $M$, the bino mass $m_{\bino}$, and the soft mass $m_{\rm soft}$. To further explore the allowed parameter space, we identify the region in the three-dimensional parameter space $\left(M, m_{\bino}, m_{\rm soft}\right)$ where the bino DM density is within the WMAP $95\%$ CL region \cite{Dunkley:2008ie}, namely $0.0975 \leq \Omega_{{\rm DM}} h^2 \leq 0.1223$. In what follows, we plot slices of this three-dimensional parameter space obtained by fixing one parameter at a constant value. 

In \Figs{fig:BoltzunifA}{fig:BoltzunifB}, we fix the soft mass to a constant value, and we identify the region in the $\left(M, m_{\bino}\right)$ plane which gives the observed DM density.  In \Fig{fig:BoltzunifA} we consider the case $m_{\rm soft} = 200~\GeV$ as in benchmarks A and B.  We immediately observe that a heavier bino requires a lighter superpotential mass. This makes sense, since we require a smaller comoving relic density $Y_{\scalars}(\infty)$ for  a heavier bino to get the same mass density, and the semi-analytical solution shows that $Y_{\scalars}(\infty)$ increases as $M$ does.  In \Fig{fig:BoltzunifB}, we consider a higher value of the soft mass, $m_{\rm soft} = 350~\GeV$, as in benchmarks C and D.  As we discussed above, a higher soft mass leads to earlier freezeout, and thus a higher DM density.  We checked that for $m_{\rm soft} \geq 400~\GeV$, the DM produced is always overabundant for the particular model in \Sec{eq:lagrangian}, but this $400~\GeV$ limit can be evaded for some of the generalizations in \Sec{sec:generalizations}. 

In \Figs{fig:BoltzunifAbis}{fig:BoltzunifBbis}, we consider slices of constant bino mass, $m_{\bino} = 180~\GeV$ and $m_{\bino} = 390~\GeV$, respectively.  We explore the allowed region in the $\left(M, m_{\rm soft}\right)$ plane.  One interesting feature is that for certain values of $m_{\bino}$ and $M$, there are two options for $m_{\rm soft}$ that reproduce the observed DM density.  For lower soft masses, freezeout occurs when $Y_{\bino} \ll Y_{\scalars}$, as for the semi-analytical case.  For a larger soft mass, freezeout occurs when $Y_{\bino} \gg Y_{\scalars}$, and we have competition between assimilation and destruction/conversion.

\section{Implications for Experiments}
\label{sec:exp}

From the point of view of DM direct and indirect detection experiments, our model has the same phenomenology as the MSSM with (nearly) pure bino DM.  Crucially, the relic density is different from the naive thermal expectation, since with assimilation, the bino abundance can be adjusted to the observed value by choosing the mass $M$ appropriately.   We saw in \Sec{sec:results} that $M$ had to be around $1~\TeV$ for assimilation to be effective, and this mass range implies spectacular collider signals.  Long-lived charged particles can be observed at the LHC, and if they are stopped, they exhibit distinctive multi-body decays.

\subsection{Dark Matter Searches}

For both direct and indirect detection experiments, the absence of any wino/higgsino content implies that the expected signals for pure bino DM are quite weak.  Generic SUSY neutralino elastic cross sections for scattering off target nuclei was computed in several references (see e.g.~\Refs{Griest:1988ma,Drees:1993bu}), for both spin-independent and spin-dependent contributions.  The primary feature of the pure bino case is the absence of $t$-channel Higgs or $Z$ exchange, which dramatically reduces the cross sections.  

The remaining contributions are from $s$-channel squark exchange.  If we assume small $A$-terms (which is natural for the first- and second-generation quarks), then there is no mixing between the left- and right-handed squarks, and the spin-independent cross section from squark exchange is highly suppressed, as can be seen from the limiting expression in \Ref{Falk:1999mq}. Thus, in a generic part of parameter space, the dominant spin-independent contribution comes from the small higgsino component, which allows for $t$-channel Higgs exchange, with a cross section \cite{Falk:1999mq,Drees:1992am}
\be
\sigma_{\bino/{\rm proton} , \, {\rm SI}} \simeq 1.76 \times 10^{-9}~\pb~\left(\frac{\tan\beta}{5}\right)^2 \left(\frac{150~\GeV}{m_{h^0}}\right)^4 \left(\frac{1~\TeV}{\mu}\right)^2 \ ,
\ee
where $\mu$ is the Higgsino mass parameter. This cross section is just below current experimental limits ($\simeq 10^{-8}~\pb$) for small values of the bino/higgsino mixing angle.  There is a spin-dependent contribution from $s$-channel squark exchange, which is well-approximated by \cite{Falk:1999mq}
\be
\sigma_{\bino/{\rm proton} , \, {\rm SD}} \simeq 7 \times 10^{-11}~\pb \, \left(\frac{1~\TeV}{m_{\tilde{q}}}\right)^4 \ ,
\ee
but this is well below current limits ($\simeq 10^{-2}~\pb$).  

The situation with indirect detection is quite pessimistic, since bino pair annihilation in our galaxy today is strongly suppressed.  The only available channel is $p$-wave suppressed annihilation to leptons and quarks, computed in \Refs{Silk:1984zy,Griest:1988ma,Drees:1992am}.  In the pure bino limit, the annihilation is usually dominated by leptons in the final state, because among the sfermions, the right-handed slepton is typically the lightest (assuming sfermion universality at a high scale) and has the highest hypercharge. The annihilation cross section is \cite{ArkaniHamed:2006mb}
\be
\sigma_{\bino \bino \; \rightarrow \; \tilde{e}_R \tilde{e}_R} \, v_{{\rm rel}} = 
4 \pi \alpha_Y^2 \frac{m_{\bino}^2 (m_{\tilde{e}_R}^4 + m_{\bino}^4)}{(m^2_{\tilde{e}_R} + m^2_{\bino})^4} \, v_{{\rm rel}}^2 \simeq  3.3 \times 10^{-8}~\pb~\left(\frac{1~\TeV}{m_{\tilde{e}_R}}\right)^4  \left(\frac{m_{\bino}}{200~\GeV}\right)^2 \ ,
\ee
which is well below current sensitivities ($\simeq 1~\pb$).

\subsection{Quasi-Stable Charged Particles at Colliders}

Unlike the feeble signals predicted in DM searches, the heavy states provide spectacular collider phenomenology.  From the relic abundance calculation, we know that $M \simeq \mathcal{O}(\TeV)$, so both $\QDirac$ and $\scalars$ can be pair produced at the LHC.  Since they are stable on collider time scales, they would be detectable as heavy stable charged particles \cite{Fairbairn:2006gg}.

\begin{figure}
\begin{center}
\includegraphics[scale=0.8]{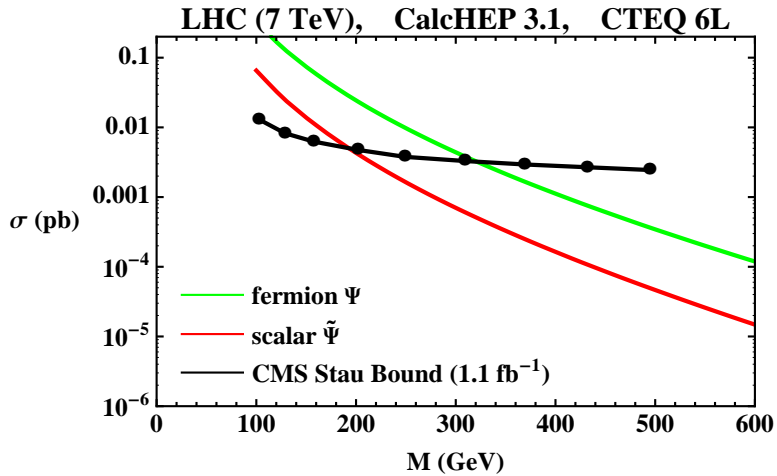}
\end{center}
\caption{Leading order cross section for pair production of fermions $\QDirac$ and scalars $\scalars$ as a function of $M$.  These cross sections are given for the 7 TeV LHC, and compared to a $1.1~\fb^{-1}$ CMS bound on quasi-stable staus \cite{Khachatryan:2011ts,CMS-PAS-EXO-11-022}.  Note that the efficiency and acceptance for the CMS search are not exactly the same as for $\QDirac$/$\scalars$ pair-production, but the bound is indicative of the LHC sensitivity.  The scalar cross section is smaller than the fermion cross section due to threshold suppression in electroweak pair production.}
\label{fig:CMSbound}
\end{figure}

Both CMS \cite{Khachatryan:2011ts,CMS-PAS-EXO-11-022} and ATLAS \cite{Aad:2011hz} already place bounds on heavy stable charged particles from anomalous ionization energy loss and time-of-flight measurements.  For the model in \Sec{eq:lagrangian}, where $\QDirac$/$\scalars$ only carry hypercharge, the most relevant study is from CMS \cite{Khachatryan:2011ts,CMS-PAS-EXO-11-022}, which bounds stable staus using the $1.1~\fb^{-1}$ data set.  In \Fig{fig:CMSbound}, we show the cross sections for the fermions $\QDirac$ and scalars $\scalars$ as a function of the mass $M$, calculated at leading order using CalcHEP 3.1 \cite{Pukhov:1999gg,Pukhov:2004ca} and CTEQ 6L parton distribution functions \cite{Pumplin:2002vw}.  The CMS study assumed that the dominant production of staus came through SUSY cascade decays, whereas in our scenario, $\QDirac$/$\scalars$ can only be pair produced through electroweak processes.   This affects somewhat the efficiency and acceptance for the search, but with that caveat in mind, $M \gtrsim 300~\GeV$ is allowed at present.

The generalizations in \Sec{sec:generalizations} are potentially more interesting, especially for colored $\Q/\Qc$ which form $R$-hadrons \cite{Farrar:1978xj}.  Because of the larger QCD pair production cross section, the LHC reach is extended to higher values of $M$.  That said, the value of $M$ typically has to be larger for colored $\Q/\Qc$, since there are extra diagrams involving gluons/gluinos which contribute to the assimilation, destruction, and conversion processes.

Most of the time, heavy stable charged particles simply exit the detector, but if they are produced with a small enough $\beta$ factor, they can loose sufficient energy through ionization that they stop inside the detector \cite{Drees:1990yw,Arvanitaki:2005nq}.  Searches for stopped exotics are currently being performed at the LHC \cite{Chen:2009gu,Khachatryan:2010uf}.  Moreover, a recent study in \Ref{Graham:2011ah} suggests that ATLAS and CMS can measure the kinematics of the $\QDirac/\scalars$ decay products.  In the model in \Sec{eq:lagrangian}, $\QDirac$ has a three-body decay to jets while $\scalars$ has a four-body decay with three jets and one DM particle.  It would be interesting to understand the extent to which these two decays can be distinguished.  The generalizations in \Sec{sec:generalizations} would have different decay patterns, but the general expectation is two quasi-stable charged states that are nearly degenerate but have different decay rates, with the lighter state decaying to $n$ SM particles, and the heavier state decaying to $n$ SM particles plus one invisible DM particle.

Finally, beyond the phenomenology of the heavy states, this scenario shares the same phenomenological features as the MSSM with the lightest neutralino as a pure bino.  In general, robustly identifying the neutralino admixture is a challenging task \cite{ArkaniHamed:2005px}, but the absence of a nearby chargino state would give some evidence that DM is a nearly pure gauge singlet (and thus must have non-standard cosmology).

\section{Conclusion}
\label{sec:con}

Upcoming results from ground- and satellite-based DM experiments as well as results from the LHC have the potential to reveal important information about the dark sector.  Assuming a positive detection of DM, one would ideally want to test whether or not DM is a thermal relic as expected from the WIMP paradigm.  For that purpose, it is important to know whether the properties of DM measured today can be extrapolated to the early universe \cite{Cohen:2008nb}.  

In this paper, we have introduced a new process called ``assimilation'', where quasi-stable particles endowed with an asymmetry can affect the abundance of DM.  These quasi-stable particles destroy DM particles but absorb DM number, and their subsequent annihilation and decay leads to the desired DM relic density.  This mechanism allows pure singlet DM to have the correct thermal relic abundance, and opens up new possibilities to achieve pure bino thermal DM in the MSSM.

While it may be surprising that long-lived particles can have such a dramatic affect on DM, it is worth remembering that the DM number density is rather small.  The asymmetry $\eta_B \simeq 8.6 \times 10^{-11}$ is larger than the DM comoving number density today, as long as $m_{\rm DM} \gtrsim 5~\GeV$.  Thus, even small asymmetries can have a large effect on DM if appropriate interactions are present.  This model offers an interesting twist on the growing literature trying to connect DM to the baryon asymmetry \cite{Kohri:2009ka,Cohen:2010kn,Shelton:2010ta,Davoudiasl:2010am,Buckley:2010ui,McDonald:2011zz,Falkowski:2011xh,Graesser:2011wi,Iminniyaz:2011yp,MarchRussell:2011fi,Davoudiasl:2011fj,Cui:2011qe,Grin:2011nk,Grin:2011tf,McDonald:2011sv,Cirelli:2011ac,Lin:2011gj,Petraki:2011mv,Kang:2011ny}.  Here, the baryon asymmetry and symmetric DM are generated by the late-time decay of a heavy particle. The idea of temporarily storing the baryon asymmetry in a quasi-stable particle may open new directions for baryogenesis (see also \Refs{McDonald:2011zz,McDonald:2011sv,Kang:2011ny}).

Despite the fact that DM is a singlet, this scenario does leaves an imprint on DM experiments.  While pure bino DM has suppressed spin-independent scattering on nuclei, only a small higgsino component is needed to give a cross section within reach of direct detection experiments.  The situation with indirect detection is less optimistic, since the same logic that leads to the expectation of bino overabundance also implies small annihilation rates in the DM halo.  The most intriguing signature of this scenario comes at the LHC, which has the potential to observe TeV-scale quasi-stable charged particles.  In this way, DM becomes visible through assimilation.

\acknowledgments We thank Spencer Chang, Clifford Cheung, and Kathryn Zurek for helpful conversations. 
Portions of this work are based on the MIT senior thesis of L.F.  This work was supported by the U.S. Department of Energy under cooperative research agreement DE-FG02-05ER-41360.  J.T. is supported by the DOE under the Early Career research program DE-FG02-11ER-41741.  

\appendix

\section{Relic Density with an Asymmetry}
\label{app:LWasymm}

In this appendix, we show that in the presence of an asymmetry, antimatter particles are efficiently destroyed in the early universe.  This justifies the fact that we neglect antimatter in the Boltzmann equations in \Sec{sec:Boltz}.

We study a simple toy example that generalizes the Lee-Weinberg scenario.  We consider a massive particle $\psi$ with mass $m_\psi$ and assume an asymmetry $\eta_\psi$ between particles and antiparticles 
\be
\eta_\psi \equiv \frac{n_\psi - n_\psibar}{s} = Y_\psi - Y_\psibar \ .
\label{eq:LWeta}
\ee
To study an example similar to our model, we consider the particles $\psi$ to be fermions charged under the hypercharge SM gauge group. The only possible channel for particle/antiparticle annihilation is
\be
\psi \psibar \rightarrow B B \ ,
\ee
where $B$ is the hypercharge gauge boson. The Boltzmann equation describing the comoving number density evolution (see \App{app:Boltzmann} for more details) is
\be
\frac{d Y_\psi}{d x} = - \frac{\lambda_{\psi \psibar \rightarrow B B}}{x^2} \left[Y_\psi Y_\psibar - Y_\psi^{\rm eq} Y_\psibar^{\rm eq}\right] \ , \qquad 
\lambda_{\psi \psibar \rightarrow B B} \equiv \frac{s(T=m_\psi)}{H(T=m_\psi)} \langle \sigma v_{{\rm rel}}\rangle_{\psi \psibar \rightarrow B B} \ ,
\label{eq:LWBoltz}
\ee
where we have defined a dimensionless time variable $x = m_\psi /T$. 

As usual, we start with equilibrium initial conditions at early times.  Because particles and antiparticles have equal and opposite chemical potentials, the product $Y_\psi^{\rm eq} Y_\psibar^{\rm eq}$ is not affected by the asymmetry, and it is equal to
\be
Y_\psi^{\rm eq} Y_\psibar^{\rm eq} = Y^2_{\psi, 0} \ , \qquad \qquad 
Y_{\psi, 0} = \frac{g_\psi}{g_{* s}} \frac{45}{4 \pi^4} x^2 \, K_2\left[ x \right] \ ,
\label{eq:psieq}
\ee
where $g_\psi$ is the number of the internal degrees of freedom for the particle $\psi$.  Using this condition, plus the asymmetry defined in \Eq{eq:LWeta}, we find
\be
Y_\psi^{\rm eq} = \frac{\eta_\psi}{2} + \sqrt{ \frac{\eta^2_\psi}{4} + Y_{\psi, 0}^2} \ , \qquad \qquad Y_\psibar^{\rm eq} = - \frac{\eta_\psi}{2} + \sqrt{ \frac{\eta^2_\psi}{4} + Y_{\psi, 0}^2} \ .
\label{eq:psimu}
\ee

\begin{figure}
\begin{center}
\subfloat[]{\includegraphics[width=2.7in]{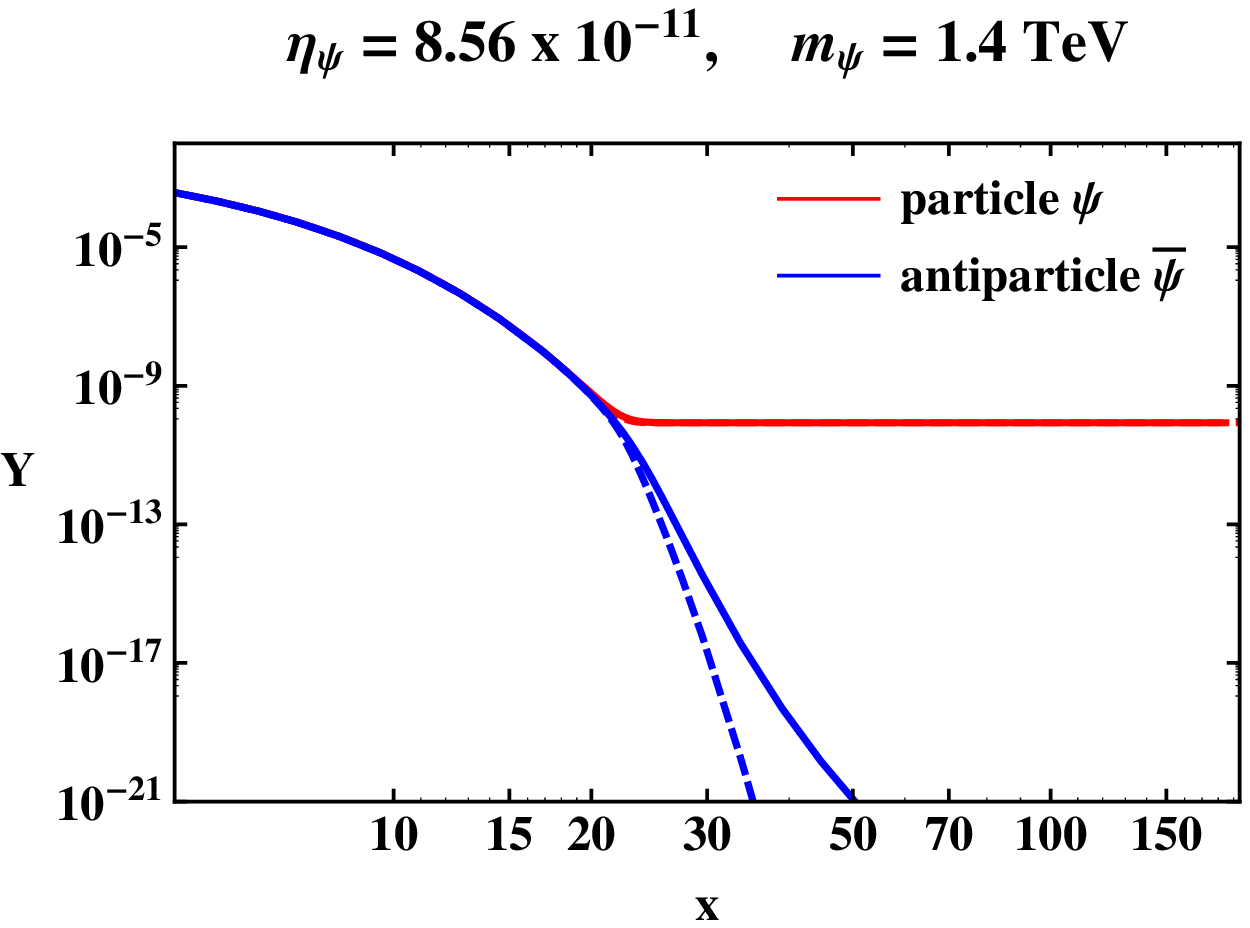}} $\qquad$
\subfloat[]{\includegraphics[width=2.7in]{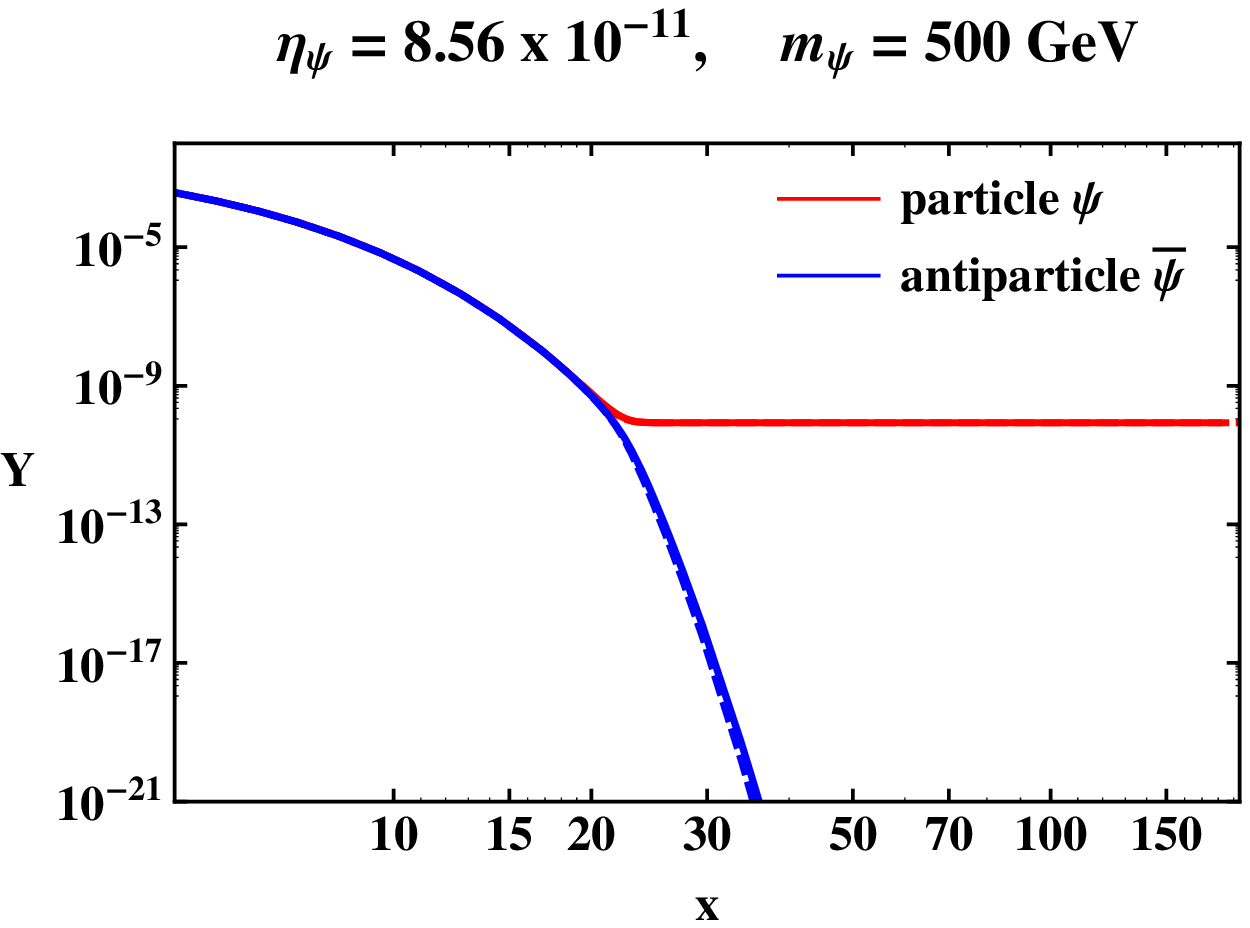}}
\end{center}
\caption{Numerical results for the Lee-Weinberg scenario with an asymmetry, fixing $\eta_\psi = 8.6 \times 10^{-11}.$ The results for the number densities of $\psi$ (red) and $\psibar$ (blue) are shown for two different particle mass: (a) $m_\psi = 1.4~\TeV$; (b) $m_\psi = 500~\GeV$. The equilibrium comoving densities are also shown in dashed lines.}
\label{fig:LWeta}
\end{figure}

The Boltzmann equation in \Eq{eq:LWBoltz} can be integrated numerically. The cross section for a $\psi \psibar$ pair to annihilate to a pair of gauge bosons is \cite{Pospelov:2007mp}
\be
\langle \sigma v_{{\rm rel}}\rangle_{\psi \psibar \rightarrow B B} = \frac{\pi \, \alpha_Y^2}{m_\psi^2} \ .
\ee
We consider two different particle masses corresponding to the benchmarks A and B defined in \Tab{tab:benchmarks}, namely $m_\psi = 1.4~\TeV$ and $m_\psi = 500~\GeV$, and motivated by the baryon asymmetry we choose $\eta_\psi \simeq 8.6 \times 10^{-11}$.  In \Fig{fig:LWeta}, we plot the comoving number densities for both $\psi$ and $\psibar$, as well as their equilibrium values.  For small values of $m_\psi$, or equivalently a larger annihilation cross section $\langle \sigma v_{{\rm rel}}\rangle_{\psi \psibar \rightarrow B B}$, the antiparticle closely tracks the equilibrium distribution in \Eq{eq:psieq} and is negligible after $x \simeq 25$. For larger value of $m_\psi$, there is an early departure from equilibrium, but we see that even for $m_\psi = 1.4~\TeV$ the full picture survives: the antiparticles are destroyed efficiently after $x \simeq 25$, and we can neglect them in our analysis.

\section{Details of the Boltzmann System}
\label{app:Boltzmann}

In this appendix, we derive the Boltzmann equations in \Eqs{eq:Boltz3eqsA}{eq:Boltz3eqsB}.  We start from a quick review of the Boltzmann equations in an expanding universe.\footnote{A detailed derivation can be found in App.~A of \Ref{D'Eramo:2010ep}.}  We consider a generic massive particle $\psi_a$ with mass $m_a$,  and we define a dimensionless time variable $x$, a comoving number density $Y_a$, and dimensionless cross sections as follows
\be
x \equiv \frac{m_a}{T} \ , \qquad Y_a \equiv \frac{n_a}{s} \ , \qquad \lambda_{ab \rightarrow cd} \equiv \frac{s(T=m_a)}{H(T=m_a)} \langle \sigma v_{{\rm rel}}\rangle_{ab \rightarrow cd} \ ,
\ee
where $T$ is the temperature, $s$ is the entropy density of the relativistic degrees of freedom, and $H$ is the Hubble parameter.  The quantity $ \langle \sigma v_{{\rm rel}}\rangle_{ab \rightarrow cd}$ denotes the thermally averaged cross section, which is constant in the $s$-wave limit. The Boltzmann equation for $Y_a$ is
\be
\begin{split}
\frac{d Y_{a}}{d x} = & \, \sum_{i} \mathcal{O}_i^{a} \ , \\ 
\mathcal{O}^{a}_{ab \rightarrow cd} = & \, -  \frac{\lambda_{ab \rightarrow cd}}{x^2}  \left[Y_a Y_b - \frac{\Yeq_a \Yeq_b}{\Yeq_c \Yeq_d} Y_c Y_d \right] 
= - \frac{\lambda_{cd \rightarrow ab}}{x^2}  \left[\frac{\Yeq_c \Yeq_d}{\Yeq_a \Yeq_b} Y_a Y_b - Y_c Y_d\right] \ ,
\end{split}
\ee
where the sum over $i$ runs over all possible reactions changing the number of $\psi_a$. The collision operator $\mathcal{O}_i^{a}$ for a generic reaction $ab \rightarrow cd$ is given in two equivalent forms, either one of which is valid.  It is typically more convenient to use the collision reaction with massless degrees of freedom in the final state such that the reaction is allowed at zero kinetic energy.

From the arguments in \App{app:LWasymm}, we can safely ignore antiparticles in our scenario.  That means that we have keep track of four species in our model: the bino $\bino$, the complex scalars $\QscalarOne$ and $\QscalarTwo$, and the Dirac fermion $\QDirac$.  Ignoring the $b_M$ term for simplicity, it is more convenient to work with $\Qscalar$ and $\overline{\Qcscalar}$.  These two complex scalars have the same mass, the same SM gauge quantum numbers, and the same interactions.   Thus, if we assume the scalars have the same initial conditions, they will always have the same number density.  For this reason we prefer to write everything in terms of the scalar comoving density $Y_{\scalars}$, defined as
\be
Y_{\scalars} \equiv Y_{\Qscalar} + Y_{\overline{\Qcscalar}} \ , \qquad \qquad Y_{\Qscalar} = Y_{\overline{\Qcscalar}} \ .
\ee

We now detail the possible reactions in our scenario.  We start from the reactions changing the bino number density
\be
\begin{split}
&
\bino \, \QDirac  \; \rightarrow \; \Qscalar \, B \ , \qquad 
\bino \, \QDirac  \; \rightarrow \; \overline{\Qcscalar}  \, B \ , \qquad
\bino \, \Qscalar \; \rightarrow \; \QDirac \, B \ , \qquad 
\bino \, \overline{\Qcscalar}  \; \rightarrow \; \QDirac \, B  \ .
\end{split}
\ee
The above reactions have the same cross sections when we replace $\overline{\Qcscalar} \leftrightarrow \Qscalar$, thus we have the following Boltzmann equation for the bino
\be
\frac{d Y_{\bino}}{d x} =   \sum_{i} \mathcal{O}_i^{\bino} = 
- 2 \frac{\lambda_{\bino \QDirac \; \rightarrow \; \Qscalar B}}{x^2} \left[Y_{\bino} Y_{\QDirac} - \frac{\Yeq_{\bino} \Yeq_{\QDirac}}{\Yeq_{\Qscalar} } Y_{\Qscalar}  \right] 
- 2 \frac{\lambda_{\bino \, \Qscalar \; \rightarrow \; \QDirac B}}{x^2} \left[Y_{\bino} Y_{\Qscalar} - \frac{\Yeq_{\bino} \Yeq_{\Qscalar}}{\Yeq_{\QDirac}} Y_{\QDirac} \right] 
 \ .
\ee
Once we express everything in terms of $Y_{\scalars} = 2 Y_{\Qscalar}$, we recover the Boltzmann equation in \Eq{eq:Boltz3eqsA}.

We now consider the scalar degrees of freedom $Y_{\scalars}$.  We first derive the Boltzmann equation for the scalar $\Qscalar$, then we replace $Y_{\scalars} = 2 Y_{\Qscalar}$. The reactions changing the number of $\Qscalar$ are
\be
\begin{split}
&
\Qscalar \, B \; \rightarrow \; \QDirac \, \bino  \ ,  \qquad 
\Qscalar \, \bino  \, \rightarrow \, \QDirac \, B \ , \qquad 
\Qscalar \, \Qscalar \; \rightarrow \; \QDirac \, \QDirac \ , \\ & 
\Qscalar \, \overline{\Qcscalar}  \; \rightarrow \; \QDirac \, \QDirac  \ ,  \qquad
\Qscalar \, \bino  \, \rightarrow \, \overline{\Qcscalar} \, \bino \ , \qquad 
\Qscalar \, \QDirac \; \rightarrow \; \overline{\Qcscalar} \, \QDirac \ , 
\end{split}
\ee
and we find the following Boltzmann equation for $Y_{\Qscalar}$
\be
\begin{split}
\frac{d Y_{\Qscalar}}{d x} =  \sum_{i} \mathcal{O}_i^{\Qscalar} = & \,
- \frac{\lambda_{\bino \QDirac \; \rightarrow \; \Qscalar B} }{x^2} \left[\frac{\Yeq_{\bino} \Yeq_{\QDirac}}{\Yeq_{\Qscalar} } Y_{\Qscalar} - Y_{\bino} Y_{\QDirac}  \right]  
- \frac{\lambda_{\bino \, \Qscalar \; \rightarrow \; \QDirac B} }{x^2} 
 \left[Y_{\bino} Y_{\Qscalar} - \frac{\Yeq_{\bino} \Yeq_{\Qscalar}}{\Yeq_{\QDirac}} Y_{\QDirac} \right] 
 \\ &  
- \frac{\lambda_{\Qscalar \Qscalar \; \rightarrow \; \QDirac\QDirac} + \lambda_{\Qscalar \overline{\Qcscalar} \; \rightarrow \; \QDirac\QDirac}  }{x^2} 
 \left[Y^2_{\Qscalar}  - \left(\frac{\Yeq_{\Qscalar}}{\Yeq_{\QDirac}}\right)^2 Y^2_{\QDirac} \right]\ .
\end{split}
\ee
With the replacement $Y_{\scalars} = 2 Y_{\Qscalar}$, we find the Boltzmann equation in \Eq{eq:Boltz3eqsB}.

We finally derive the Boltzmann equation for the Dirac fermion $\QDirac$, which is not given in the main text.  The reactions changing its number density are
\be
\begin{split}
&
\QDirac \, \bino  \; \rightarrow \; \Qscalar \, B \ , \qquad 
\QDirac \, \bino \; \rightarrow \; \overline{\Qcscalar} \, B \ , \qquad
\QDirac \, B  \; \rightarrow \; \Qscalar \, \bino  \ , \qquad 
\QDirac \, B \; \rightarrow \;  \overline{\Qcscalar} \, \bino  \ , \\ &
\QDirac \, \QDirac \; \rightarrow \; \Qscalar \, \Qscalar \ , \qquad 
\QDirac \, \QDirac \; \rightarrow \; \overline{\Qcscalar} \, \overline{\Qcscalar} \ , \qquad 
\QDirac \, \QDirac \; \rightarrow \; \Qscalar \, \overline{\Qcscalar}  \ ,
\end{split}
\ee
with corresponding Boltzmann equation
\begin{align}
\frac{d Y_{\QDirac}}{d x} =   \sum_{i} \mathcal{O}_i^{\QDirac} = & \,
- 2 \frac{\lambda_{\bino \QDirac \; \rightarrow \; \Qscalar B}}{x^2} 
\left[Y_{\bino} Y_{\QDirac} - \frac{\Yeq_{\bino} \Yeq_{\QDirac}}{\Yeq_{\Qscalar} } Y_{\Qscalar}  \right] 
 - 2 \frac{\lambda_{\bino \, \Qscalar \; \rightarrow \; \QDirac B}}{x^2} 
 \left[\frac{\Yeq_{\bino} \Yeq_{\Qscalar}}{\Yeq_{\QDirac}} Y_{\QDirac} - Y_{\bino} Y_{\Qscalar} \right]   \nonumber
\\ &
-  \frac{\lambda_{\Qscalar \Qscalar \; \rightarrow \; \QDirac\QDirac} + \lambda_{\Qscalar \overline{\Qcscalar} \; \rightarrow \; \QDirac\QDirac}}{x^2} 
\left[ \left(\frac{\Yeq_{\Qscalar}}{\Yeq_{\QDirac}}\right)^2 Y^2_{\QDirac} - Y^2_{\Qscalar} \right] \ .
\end{align}
As a cross check, we observe that the sum of the $\QDirac$ and $\scalars$ Boltzmann equations gives
\be
\frac{d}{dx} \left[ Y_{\scalars} + Y_{\QDirac} \right] = 0 \qquad \Rightarrow \qquad Y_{\scalars} + Y_{\QDirac} = {\rm const} = \etaB \ ,
\ee
consistent with conservation of $U(1)_B$.  Hence, we only need to solve a system of two Boltzmann equations, and we choose to solve for $\bino$ and $\scalars$.

\section{Interactions and Thermal Cross Sections}
\label{app:amplitudes}

In this appendix, we give the interactions and thermal cross section for the model in \Sec{eq:lagrangian}.  For simplicity, we consider the $s$-wave limit of the various interactions.

The components of $\Q$ and $\Qc$ couple to the SM hypercharge gauge boson as
\be
\mathcal{L} \supset \left|D^\mu \Qscalar \right|^2 + \left|D^\mu \Qcscalar\right|^2 
+ i \Qfermiondag \sigmabar^\mu D_\mu \Qfermion + i \Qcfermiondag \sigmabar^\mu D_\mu \Qcfermion \ ,
\ee
where the covariant derivatives are
\be
D_\mu \Qscalar = \partial_\mu \Qscalar + i  g_Y B_\mu \Qscalar  \ , \qquad  
D_\mu \Qcscalar = \partial_\mu \Qcscalar - i  g_Y B_\mu \Qcscalar  \ ,
\ee
\be
D_\mu \Qfermion = \partial_\mu \Qfermion + i g_Y B_\mu \Qfermion  \ , \qquad 
D_\mu \Qcfermion = \partial_\mu \Qcfermion - i g_Y B_\mu \Qcfermion \ .
\ee
Likewise, they couple to the bino via
\be
\mathcal{L} \supset
- \sqrt{2} g_Y \left[ \left(\Qscalar{\,^\dag} \, \Qfermion - \Qcscalarstar \, \Qcfermion \right) \; \bino + \overline{\bino} \left( \Qfermiondag \, \Qscalar - \Qcfermiondag \, \Qcscalar \right)\right]  \ .
\ee 
Once we integrate out the auxiliary $D$ fields, we also get a scalar potential
\be
V \left(\Qscalar, \Qcscalar\right) = \frac{1}{2} g_Y^2 \left[ |\Qscalar|^2 - |\Qcscalar|^2 + \text{MSSM fields} \right]^2 \ ,
\label{eq:VSUSY}
\ee 
which is not relevant for our discussion.

From the interactions above, it is straightforward to compute the thermally averaged cross sections $\langle \sigma v_{{\rm rel}} \rangle$ and use them to solve the Boltzmann equation system.  We implement our model in FeynRules \cite{Christensen:2008py}, using the superspace module described in \Ref{Duhr:2011se}.  We then use the FeynRules interface to translate the model to CalcHEP format \cite{Pukhov:1999gg,Pukhov:2004ca}, which we use to analytically compute the squared amplitudes $\left|\mathcal{M}_{a b \rightarrow c d}\right|^2$.  In the $\delta \ll M$ limit, the squared matrix elements averaged over the initial and summed over the final polarizations are
\be
\begin{split}
\left|\mathcal{M}_{\bino \QDirac \; \rightarrow \; \Qscalar B}\right|^2  & =
g_Y^4 \frac{8 M^4+4 M^3 m_{\bino}-10 M^2 m_{\bino}^2-4 M m_{\bino}^3+m_{\bino}^4}{M m_{\bino} (2 M+m_{\bino})^2} + \mathcal{O}\left(\frac{\delta}{M}\right)\ , \\ 
\left|\mathcal{M}_{\bino \, \Qscalar \; \rightarrow \; \QDirac B}\right|^2  & =
2 g_Y^4  \frac{8 M^4+20 M^3 m_{\bino}+18 M^2 m_{\bino}^2+6 M m_{\bino}^3+m_{\bino}^4}{m_{\bino} (M+m_{\bino}) (2 M+m_{\bino})^2} + \mathcal{O}\left(\frac{\delta}{M}\right) \ , \\ 
\left|\mathcal{M}_{\Qscalar \Qscalar \; \rightarrow \; \QDirac\QDirac}\right|^2  & = 16 g_Y^4  \frac{M^2}{m_{\bino}^2} + \mathcal{O}\left(\frac{\delta}{M}\right) \ , \\ 
\left|\mathcal{M}_{\Qscalar \overline{\Qcscalar} \; \rightarrow \; \QDirac\QDirac}\right|^2  & =  32 g_Y^4 \frac{M^3}{m_{\bino}^4} \delta + \mathcal{O}\left(\frac{\delta^2}{M^2}\right) \ .
\end{split}
\label{eq:Msquared}
\ee
For a generic process $a b \rightarrow c d$ in the $s$-wave limit, we have\footnote{A derivation of this result can be found, for example, in App.~C of \Ref{D'Eramo:2010ep}.}
\be
\langle \sigma v_{{\rm rel}}\rangle_{a b \rightarrow c d} = \frac{\left|\mathcal{M}_{a b \rightarrow c d}\right|^2}{32 \pi \, m_a m_b} 
\left[1 - 2 \frac{(m_c^2+m_d^2)}{(m_a + m_b)^2} + \frac{(m_c^2-m_d^2)^2}{(m_a + m_b)^4}\right]^{1/2} \ .
\ee
The squared matrix element $\left|\mathcal{M}_{a b \rightarrow c d}\right|^2$ is function of the Mandelstam variables $s$ and $t$, and for the $s$-wave amplitudes we have
\be
s \simeq \left(m_a + m_b\right)^2 \ , \qquad \qquad t \simeq \frac{m_b m_c^2 + m_a m_d^2}{m_a + m_b} - m_a m_b \ .
\ee


\bibliographystyle{JHEP}
\bibliography{DMAssimilation}

\end{document}